\documentclass[aps,prd,twocolumn,superscriptaddress]{revtex4-1}

\usepackage{graphicx}
\usepackage{latexsym}
\usepackage{bm}
\usepackage{amsmath}
\usepackage{amssymb}
\usepackage{ifthen}

\renewcommand{\eqref}[1]{(\ref{#1})}

\newcommand{\kf}{(k_{F})}
\newcommand{\wxyz}[4]{w^{#1}x^{#2}y^{#3}z^{#4}}
\newcommand{\diag}{\ensuremath{\text{\diag}}}

\newcommand{\ktr}{\tilde{\kappa}_{tr}}
\newcommand{\kem}{\tilde{\kappa}_{e-}}
\newcommand{\kep}{\tilde{\kappa}_{e+}}
\newcommand{\kop}{\tilde{\kappa}_{o+}}
\newcommand{\kom}{\tilde{\kappa}_{o-}}

\newcommand{\epv}[2]{\ifthenelse{\equal{#2}{-k}}{\ensuremath{\vec{\epsilon}_{#1}(-\vec{k})}}{\ensuremath{\vec{\epsilon}_{#1}(\vec{#2})}}}
\newcommand{\kemktr}{\left(\kem+I\ktr\right)}
\newcommand{\epvkemktrepv}[3]{\left[\epv{#1}{#3}\cdot\kemktr\cdot\epv{#2}{#3}\right]}
\newcommand{\epvkopepv}[3]{\left[\epv{#1}{#3}\cdot\kop\cdot\epv{#2}{#3}\right]}
\newcommand{\wtcal}[1]{\widetilde{\mathcal{#1}}}

\newcommand{\kA}{\mathcal{A}}
\newcommand{\kAu}[2]{\ifthenelse{\equal{#2}{-k}}{\ensuremath{\mathcal{A}^{#1}(-\vec{k})}}{\ensuremath{\mathcal{A}^{#1}(\vec{#2})}}}
\newcommand{\kAd}[2]{\ifthenelse{\equal{#2}{-k}}{\ensuremath{\mathcal{A}^{#1}(-\vec{k})}}{\ensuremath{\mathcal{A}_{#1}(\vec{#2})}}}
\newcommand{\kpu}[2]{\ifthenelse{\equal{#2}{-k}}{\ensuremath{\pi^{#1}(-\vec{k})}}{\ensuremath{\pi^{#1}(\vec{#2})}}}
\newcommand{\kpd}[2]{\ifthenelse{\equal{#2}{-k}}{\ensuremath{\pi_{#1}(-\vec{k})}}{\ensuremath{\pi_{#1}(\vec{#2})}}}

\newcommand{\ket}[1]{|#1\rangle}
\newcommand{\bra}[1]{\langle #1|}
\newcommand{\iket}[1]{|#1\!\!\supset}
\newcommand{\ibra}[1]{\subset\!\!#1|}

\newcommand{\ibraket}[2]{\ibra{#1}#2\!\!\supset}

\renewcommand{\ao}[2]{\ifthenelse{\equal{#2}{}}{\ensuremath{a_{#1}}}{\ifthenelse{\equal{#2}{-k}}{\ensuremath{a_{#1}(-\vec{k})}}{\ensuremath{a_{#1}(\vec{#2})}}}}
\newcommand{\ab}[2]{\ifthenelse{\equal{#2}{}}{\ensuremath{\overline{a}_{#1}}}{\ifthenelse{\equal{#2}{-k}}{\ensuremath{\overline{a}_{#1}(-\vec{k})}}{\ensuremath{\overline{a}_{#1}(\vec{#2})}}}}
\newcommand{\ad}[2]{\ifthenelse{\equal{#2}{}}{\ensuremath{a^{\dagger}_{#1}}}{\ifthenelse{\equal{#2}{-k}}{\ensuremath{a^{\dagger}_{#1}(-\vec{k})}}{\ensuremath{a^{\dagger}_{#1}(\vec{#2})}}}}
\newcommand{\abd}[2]{\ifthenelse{\equal{#2}{}}{\ensuremath{\overline{a}^{\dagger}_{#1}}}{\ifthenelse{\equal{#2}{-k}}{\ensuremath{\overline{a}^{\dagger}_{#1}(-\vec{k})}}{\ensuremath{\overline{a}^{\dagger}_{#1}(\vec{#2})}}}}

\newcommand{\bo}[2]{\ifthenelse{\equal{#2}{}}{\ensuremath{b_{#1}}}{\ifthenelse{\equal{#2}{-k}}{\ensuremath{b_{#1}(-\vec{k})}}{\ensuremath{b_{#1}(\vec{#2})}}}}
\newcommand{\bb}[2]{\ifthenelse{\equal{#2}{}}{\ensuremath{\overline{b}_{#1}}}{\ifthenelse{\equal{#2}{-k}}{\ensuremath{\overline{b}_{#1}(-\vec{k})}}{\ensuremath{\overline{b}_{#1}(\vec{#2})}}}}
\newcommand{\bd}[2]{\ifthenelse{\equal{#2}{}}{\ensuremath{b^{\dagger}_{#1}}}{\ifthenelse{\equal{#2}{-k}}{\ensuremath{b^{\dagger}_{#1}(-\vec{k})}}{\ensuremath{b^{\dagger}_{#1}(\vec{#2})}}}}
\newcommand{\bbd}[2]{\ifthenelse{\equal{#2}{}}{\ensuremath{\overline{b}^{\dagger}_{#1}}}{\ifthenelse{\equal{#2}{-k}}{\ensuremath{\overline{b}^{\dagger}_{#1}(-\vec{k})}}{\ensuremath{\overline{b}^{\dagger}_{#1}(\vec{#2})}}}}

\newcommand{\adn}[2]{(a^{\dagger}_{#1})^{#2}}

\newcommand{\ie}{{\it i.e.\ }}
\newcommand{\eg}{{\it e.g.\ }}

\begin{document}

\title{Covariant Quantization of Lorentz-Violating Electromagnetism}


\author{Michael A. Hohensee}
\affiliation{Harvard-Smithsonian Center for Astrophysics, Cambridge, Massachusetts, 02138, USA}
\affiliation{Department of Physics, Harvard University, Cambridge, Massachusetts, 02138, USA}
\affiliation{Department of Physics, University of California, Berkeley, 94720, USA}
\author{David F. Phillips}
\affiliation{Harvard-Smithsonian Center for Astrophysics, Cambridge, Massachusetts, 02138, USA}
\author{Ronald L. Walsworth}
\affiliation{Harvard-Smithsonian Center for Astrophysics, Cambridge, Massachusetts, 02138, USA}
\affiliation{Department of Physics, Harvard University, Cambridge, Massachusetts, 02138, USA}

\date{\today}

\begin{abstract}
We present a consistent, generally covariant quantization of light for non-vacuum birefringent, Lorentz-symmetry breaking electrodynamics in the context of the Standard Model Extension.  We find that the number of light quanta in the field is not frame independent, and that the interaction of the quantized field with matter is necessarily birefringent.  We also show that the conventional Lorenz gauge condition used to restrict the photon-mode basis to solutions of the Maxwell equations must be weakened to consistently describe Lorentz symmetry violation.
\end{abstract}


\maketitle

The Maxwell equations are invariant under arbitrary Lorentz transformations, and thus the speed of light is constant and isotropic in all reference frames.  This statement is a cornerstone of modern physics, and as such, has been subject to a wide variety of experimental tests of ever increasing precision for more than a century~\cite{Michelson,IvesStillwell,KennedyThorndike,Hall,PetersGroup,Tobar:2005,TobarGroup,ModernIvesStillwell,Collider,ComptonEdge}.  More recent work has focused on using tests of Lorentz invariance to search for the low-energy imprint, \eg spontaneous Lorentz symmetry breaking, of physics at higher energy scales~\cite{Kostelecky:1989,Colladay:1997,Colladay:1998,Mattingly:2005}. Today, many such tests are rigorously analyzed and compared to one another using the standard model extension (SME)~\cite{Colladay:1997,Colladay:1998,KosteleckyRussellRMP}, an effective field theory that includes all of the standard model of particle physics as a limiting case, and augments it with all Lorentz-scalar operators that can be constructed from standard model fields that are not term-by-term invariant under Lorentz transformation.  Here, we are primarily concerned with the quantized representation of the free electromagnetic field in the context of the minimal SME, which includes only operators of mass dimension $3$ or $4$, and specifically focus on those operators not already subject to stringent observational constraints from astrophysics.

Most experimental and theoretical investigations of Lorentz-violating electrodynamics to date have treated the fields classically, as in analyses of Michelson-Morley tests~\cite{PetersGroup,TobarGroup}, or semiclassically with the assumption that the excitations of the quantized fields satisfy the classical dispersion relation, as in Ives-Stilwell experiments~\cite{ModernIvesStillwell}.  In situations for which a fully quantum treatment of both vacuum-non-birefringent electromagnetism and the coupled charges is necessary, quantization is formally preceded by a coordinate redefinition which maps the anisotropy in the speed of light into an anisotropy in the maximum attainable speed of all other particles.  Quantized theories of such Lorentz-symmetry breaking matter have been demonstrated to be stable~\cite{Colladay:1997,Colladay:1998}.  In effect, the anisotropy the electromagnetic sector is masked by using the wavelength of a photon of fixed frequency as a rod to measure distance.  This step adds complexity to many theoretical analyses of a given Lorentz symmetry test, and may in some cases obscure some of the interesting features of both the Lorentz-violating and fully covariant theory.  Worse, this added complexity may sometimes lead researchers to begin with the arbitrary assumption that one or more sectors of the theory are exactly Lorentz-invariant, greatly complicating efforts to make rigorous global comparisons of results between different experiments.

Here, we make some initial steps towards deriving a fully general, quantized Hamiltonian representation of electrodynamics in the photon sector of the SME, focusing on the quantization of the freely propagating field.  We demonstrate that the Hamiltonian that results from the photon-sector Lagrangian is Hermitian, and so does not violate unitarity.  Furthermore, we show that the quantized Hamiltonian leaves the subspace of states corresponding to solutions of the Lorentz-violating Maxwell equations invariant. We find that the quantized modes reproduce the dispersion relation obtained from the classical Lorentz-violating theory \cite{Colladay:1998}.  

In part~\ref{sec:smephoton}, we briefly review the photon-sector of the minimal SME, defining approximations and notation that will be used throughout the rest of our analysis.  In part~\ref{sec:lorentzfermi}, we define the Lorentz-symmetry breaking Fermi Lagrangian, and the form of the conjugate field momenta $\pi_{\gamma}(\vec{k})$ to first order in the Lorentz-violating interaction.

In part~\ref{sec:Hamiltonian}, we obtain the explicit form of the Lorentz-violating Hamiltonian operator, and quantize the theory.  We also obtain the unitary transformation that diagonalizes the Hamiltonian operator in terms of the normal modes of the fully covariant theory.  This transformation is frame-dependent, and is consistent with the observation made in \cite{Mattingly:2005} that the vacuum apparent in one inertial frame may not be equivalent to that in other frames, as happens, \eg when comparing the vacuum of the covariant theory in an inertial frame with that in an accelerated frame \cite{Fulling:1973,Fuentes-Schuller:2005a}.  This work may provide a path to apply recent developments in relativistic quantum information to tests of Lorentz invariance~\cite{Peres:2004}.  In part~\ref{sec:whyindef}, we consider the form of the theory in the indefinite metric imposed by our choice of covariant commutator, and show that the Hamiltonian derived in part~\ref{sec:Hamiltonian} is hermitian, and that the evolution of the quantized Lorentz-violating theory is unitary.  We find that the Faddeev-Popov ghost degrees of freedom~\cite{Kugo:1978} are necessary to the development of the Lorentz-violating theory.  Although they do not themselves contribute to physical observables, they do provide a channel for vacuum-birefringent effects at second order in the Lorentz-symmetry breaking parameters.  Thus in part~\ref{subsec:weaklorenz}, we apply a subsidiary gauge condition that is weaker than that used by Gupta and Bleuler~\cite{GuptaBleuler} in their quantization of the fully covariant theory.


We close by briefly considering the form of the transverse potentials in terms of the free-field eigenmode operators in part \ref{sec:quanthameffects}.  The unitary transformation derived at the end of \ref{sec:Hamiltonian} is shown to lead to anisotropic scaling as well as mixing between the transverse potentials.  This suggests that the ``non-birefringent'' components of $\kf$ could lead to a polarization-dependent, and thus birefringent coupling between light and an isotropic medium it passes through, consistent with recent analyses of the classical~\cite{Kostelecky:2009}, and the coordinate-transformed semi-classical~\cite{Mueller:2005} theory.  

\section{The Photon Sector of the SME\label{sec:smephoton}}
In the photon sector of the minimal SME, the conventional $-\frac{1}{4}F^{2}$ electromagnetic Lagrangian is augmented to become~\cite{Colladay:1997,Colladay:1998,Kostelecky:2002}
\begin{equation}
\mathcal{L}=-\frac{1}{4}F_{\mu\nu}F^{\mu\nu}-\frac{1}{4}{\kf}_{\kappa\lambda\mu\nu}F^{\kappa\lambda}F^{\mu\nu}+\frac{1}{2}(k_{AF})^{\kappa}\epsilon_{\kappa\lambda\mu\nu}A^{\lambda}F^{\mu\nu},\label{eq:smefullphotonlagrange}
\end{equation}
where both $\kf$ and $(k_{AF})$ break particle Lorentz symmetry.  The $(k_{AF})$ term also breaks CPT symmetry, and has units of mass.  The best constraints upon $(k_{AF})$ are derived from polarization studies of the cosmic microwave background, and are presently such that the magnitude of each of the four components is estimated to be no larger than $\sim 10^{-43}$~GeV \cite{KosteleckyRussellRMP,Kostelecky:2008a}.  This is far below the scale at which the elements of $\kf$ have been constrained, and is indeed far below the reach of any proposed experimental investigations, which are sensitive to $(k_{AF})$ at the level of $\sim10^{-21}$~GeV~\cite{Kostelecky:2008a,Mewes:2008}.  Accordingly, we will consider only models in which $(k_{AF})=0$ in our subsequent analyses.  The $\kf$ tensor has the symmetries of the Riemann tensor and a vanishing double trace, and thus actually represents only 19 independent parameters.  The dimensionless $\kf$ does not generate a photon mass, but instead imparts fractional variations in the phase velocity of electromagnetic waves propagating in a Lorentz-symmetry violating vacuum.  These variations can depend upon the both the direction and polarization of the propagating wave.  This anisotropy can be formally removed from the photon sector at leading order by the coordinate transformation~\cite{Kostelecky:2001,Kostelecky:2002,Bailey:2004,Altschul:2009}
\begin{equation}
x'^{\mu}=x^{\mu}-\frac{1}{2}{\kf}^{\alpha\mu}_{\phantom{\alpha\mu}\alpha\nu}c^{\nu},
\end{equation}
which maps ${\kf}^{\alpha}_{\phantom{\alpha}\mu\alpha\nu}\rightarrow 0$, and generates corresponding anisotropies in the matter-sector, where quantization has already been demonstrated~\cite{Colladay:1997,Colladay:1998}.  
In \cite{Kostelecky:2002}, the $\kf$ tensor is re-expressed in the more phenomenologically transparent form as
\begin{align}\begin{split}
\mathcal{L}&=\frac{1}{2}\left[(1+\ktr)|\vec{E}|^{2}-(1-\ktr)|\vec{B}|^{2}\right]\\
&\quad\quad+\frac{1}{2}\left[\vec{E}\cdot\left(\kep+\kem\right)\cdot\vec{E}-\vec{B}\cdot\left(\kep-\kem\right)\cdot\vec{B}\right]\\
&\quad\quad\quad+\vec{E}\cdot\left(\kop+\kom\right)\cdot\vec{B},\label{eq:smeEBLag}
\end{split}\end{align}
where $\ktr$ is a scalar; and the $3\times 3$ $\kep$, $\kem$, $\kom$ matrices are traceless and symmetric, while $\kop$ is antisymmetric.  In terms of $\kf$, the $\tilde{\kappa}$'s are given by~\cite{Kostelecky:2002}
\begin{align}\begin{split}
(\kep)^{jk}&=\:-(k_{F})^{0j0k}+\frac{1}{4}\epsilon^{jpq}\epsilon^{krs}(k_{F})^{pqrs},\\
(\kem)^{jk}&=\:-(k_{F})^{0j0k}-\frac{1}{4}\epsilon^{jpq}\epsilon^{krs}(k_{F})^{pqrs}+\frac{2}{3}\delta^{jk}(k_{F})^{0l0l},\\
(\kop)^{jk}&=\:\frac{1}{2}\left((k_{F})^{0jpq}\epsilon^{kpq}-(k_{F})^{0kpq}\epsilon^{jpq}\right),\\
(\kom)^{jk}&=\:\frac{1}{2}\left((k_{F})^{0jpq}\epsilon^{kpq}+(k_{F})^{0kpq}\epsilon^{jpq}\right),\\
\ktr&=\:-\frac{2}{3}(k_{F})^{0l0l}.
\end{split}\end{align}
Sums on repeated roman indices $j,k,m,p,q,r,s=1,2,3$ are implied.  We then define the electromagnetic fields, as originally outlined in \cite{Colladay:1997,Colladay:1998} and \cite{Kostelecky:2001,Kostelecky:2002}, as
\begin{equation}
\left(\begin{matrix}\vec{D}\\ \vec{H}\end{matrix}\right)=\left(\begin{matrix} (1+\ktr)+\kep+\kem & \kop+\kom \\ \kop+\kom & (1-\ktr)+\kep-\kem\end{matrix}\right)\left(\begin{matrix}\vec{E}\\ \vec{B}\end{matrix}\right),
\end{equation}
then the Lagrangian equations of motion derivable from \eqref{eq:smeEBLag} reduce to the form of the Maxwell equations in an anisotropic medium
\begin{align}\begin{split}
\vec{\nabla}\times\vec{H}-c\,\partial_{t}\vec{D}&=0, \quad \vec{\nabla}\cdot\vec{D}=0\\
\vec{\nabla}\times\vec{E}-c\,\partial_{t}\vec{B}&=0, \quad \vec{\nabla}\cdot\vec{B}=0.
\end{split}\end{align}
This implies that the general form of the solution to the wave equation in the Lorentz-violating vacuum is similar to that of a plane wave propagating in an anisotropic medium.  We can immediately see that $\ktr$ gives rise to an isotropic shift in the effective permeability and permittivity of the vacuum, and thus an isotropic and helicity-independent shift in the speed of light \cite{Kostelecky:2002}.  To determine the effects of the other $\tilde{\kappa}$'s, we need to solve the full dispersion relation.  The analogy with electromagnetism in anisotropic media leads us to write the ansatz
\begin{align}
\vec{E}&=\vec{E}_{0}e^{-i\omega t + i\vec{k}\cdot\vec{r}}&\text{ and }&& \vec{B}&=\vec{B}_{0}e^{-i\omega t + i\vec{k}\cdot\vec{r}},
\end{align}
and require that $\omega$, $\vec{k}$, and the fields satisfy the modified Amp\`ere law \cite{Colladay:1997,Colladay:1998,Colladay:2001,Kostelecky:2001,Kostelecky:2002}
\begin{equation}
\left(-\delta^{pq}k^{2}-k^{p}k^{q}-2\kf^{p\beta\gamma q}k_{\beta}k_{\gamma}\right)E^{q}=0.\label{eq:lvampere}
\end{equation}
To leading order in $\kf$, this modifies the dispersion relation between $\omega$ and $\vec{k}$, yielding
\begin{equation}
\omega_{\pm}=(1+\rho\pm\sigma)|\vec{k}|c.\label{eq:smedispersion}
\end{equation}
The $\pm$ subscript on $\omega$ and between $\rho$ and $\sigma$ denotes whether the wave has positive or negative helicity, so that $\rho$ represents a polarization-independent shift of the phase velocity, while $\sigma$ is a birefringent shift.  In terms of $\kf$, these parameters are
\begin{align}
\rho&=-\frac{1}{2}\widetilde{k}_{\alpha}^{\phantom{\alpha}\alpha}, &\sigma^{2}&=\frac{1}{2}\widetilde{k}_{\alpha\beta}\widetilde{k}^{\alpha\beta}-\rho^{2},\label{eq:krhosigma}
\end{align}
where
\begin{align}
\widetilde{k}^{\alpha\beta}&=\kf^{\alpha\mu\beta\nu}\hat{k}_{\mu}\hat{k}_{\nu}, & \hat{k}_{\mu}&=k_{\mu}/|\vec{k}|\label{eq:kab}
\end{align}
and $k_{\mu}$ is the four-vector $(\omega/c,\vec{k})$, and the relativistic inner product is implied by pairs of repeated subscripted and superscripted greek indices: $A_{\mu}B^{\mu}=A_{0}B_{0}-A_{1}B_{1}-A_{2}B_{2}-A_{3}B_{3}$.  The $\rho$ and $\sigma$ governing the dispersion relation for a plane wave propagating in the $+\hat{z}$ direction may be written in terms of the $\tilde{\kappa}$'s as \cite{Tobar:2005}
\begin{equation}
\rho=-\ktr+\frac{1}{2}\kem^{33}+\kop^{12}
\end{equation}
and
\begin{equation}
\sigma^{2}=\frac{1}{4}\left(\kom^{11}-\kom^{22}-2\kep^{12}\right)^{2}+\frac{1}{4}\left(\kep^{22}-\kep^{11}-2\kom^{12}\right)^{2}.
\end{equation}
Note that $\ktr$, $\kop$, and $\kem$ govern the polarization-independent shifts, while $\kep$ and $\kom$ describe birefringence.  Because the theory is invariant under observer rotations, this division holds for waves propagating in any direction.  The division persists to first order in the $\tilde{\kappa}$'s under boosts of the observer frame, since observer Lorentz covariance requires that observing birefringent phenomena in one inertial frame implies birefringence in all frames, while its absence in one frame implies its absence in all other frames~\cite{note1}.

The ten birefringent parameters $\kom$ and $\kep$ components of the $\kf$ tensor have been constrained at the level of $10^{-37}$ by spectropolarimetric studies of light emitted from distant stars \cite{Kostelecky:2001,Kostelecky:2002,Kostelecky:2006}.  A comparatively weak constraint of $10^{-16}$ on the birefringent $\tilde{\kappa}$'s was obtained in \cite{Kostelecky:2002} by searching for evidence of birefringence-induced time-splitting of short pulses of light emitted from distant millisecond pulsars and gamma-ray bursts.  The far stronger constraint of $10^{-32}$ \cite{Kostelecky:2002} and even $10^{-37}$ for some combinations of $\kom$ and $\kep$ \cite{Kostelecky:2006} is derived from searches for characteristic correlations between the polarization and wavelength of light observed from distance sources.  These constraints are far stronger than the best limits on the nine non-birefringent $\ktr$, $\kop$, and $\kem$ parameters, and thus the contribution of the $\kom$ and $\kep$ matrices will be neglected in our subsequent analyses.  Taking this approximation, we may write down the fractional shift $\delta(\vec{k})$ in the vacuum phase velocity of light moving in arbitrary directions in terms of its transverse polarization vectors
\begin{align}\begin{split}
\delta(\vec{k})&=\:\epvkopepv{1}{2}{k}\\
&\:\quad-\frac{1}{2}\sum_{r=1}^{2}\epvkemktrepv{r}{r}{k},\label{eq:nonbirefringentvelocity}
\end{split}\end{align}
where for each $\vec{k}$, the transverse unit polarization vectors $\epv{1}{k}$ and $\epv{2}{k}$ satisfy
\begin{align}
\begin{split}
\epv{1}{k}=\epv{1}{-k}\quad  \epv{2}{k}&=-\epv{2}{-k},\\
\text{ and }  \quad\epv{1}{k}\times\epv{2}{k}&=\hat{k}=\epv{3}{k}=-\epv{3}{-k}.\label{eq:polarizationdefs}
\end{split}
\end{align}
As an illustrative example of the roles played by the different non-birefringent $\tilde{\kappa}$ parameters, we see that for light traveling along the $z$-axis in the $+z$ direction, with $\epv{1}{k\hat{z}}=\hat{x}$ and $\epv{2}{k\hat{z}}=\hat{y}$,
\begin{equation}
\delta(k\hat{z})=\kop^{xy}-\ktr-\frac{1}{2}\left(\kem^{xx}+\kem^{yy}\right)=\kop^{xy}-\ktr+\frac{1}{2}\kem^{zz},\label{eq:rhozplus}
\end{equation}
where we have taken advantage of the vanishing trace of $\kem$.  For light traveling in the $-z$ direction, however, we find that
\begin{equation}
\delta(-k\hat{z})=-\kop^{xy}-\ktr+\frac{1}{2}\kem^{zz},\label{eq:rhozminus}
\end{equation}
since \eqref{eq:polarizationdefs} specifies the sign of $\epv{1,2}{k}$ relative to $\epv{1,2}{-k}$.
Thus $\ktr$ represents an isotropic fractional reduction in the vacuum phase velocity of light, $\kem$ describes the average shift in the speed of light propagating back and forth along a given axis, and $\kop$ governs the difference in the one-way speed of light along an axis.

\section{The Lorentz-Violating Fermi Lagrangian\label{sec:lorentzfermi}}

We begin with the photon-sector free-field Lagrangian density
\begin{equation}
\mathcal{L}=-\frac{1}{4}F_{\mu\nu}F^{\mu\nu}-\frac{1}{4}{\kf}_{\kappa\lambda\mu\nu}F^{\kappa\lambda}F^{\mu\nu},\label{eq:smelagrangian}
\end{equation}
where $F_{\mu\nu}=\partial_{\nu}A_{\mu}-\partial_{\mu}A_{\nu}$, and we have assumed $(k_{AF})=0$ (see part \ref{sec:smephoton}).  Direct canonical quantization of the potential $A_{\mu}$ using \eqref{eq:smelagrangian} is impossible since observer Lorentz invariance requires the commutator between the quantized fields to be a Lorentz scalar, and the momentum $\pi^{0}$ conjugate to the scalar potential $A^{0}$ is given by
\begin{align}
\pi^{0}&=\frac{\partial\mathcal{L}}{\partial\dot{A}^{0}}=0.
\end{align}
This is a reflection of the fact that the scalar potential is not a physical observable.  This problem can be addressed by quantizing an observable like $\vec{E}$, in place of the physically unobservable vector potential $A^{\mu}$, but taking such a step at this stage would complicate the form of the interaction with charges, and obscure the Lorentz covariance of the $F^{2}$ component of the Lagrangian.  Our first step is therefore to find an alternative Lagrangian which produces the same physics.  The equations of motion which result from \eqref{eq:smelagrangian} are
\begin{equation}
\partial^{\alpha}\frac{\partial\mathcal{L}}{\partial(\partial^{\alpha}A^{\gamma})}=\partial^{\alpha}F_{\alpha\gamma}+\partial^{\alpha}{\kf}_{\alpha\gamma}F^{\mu\nu}=0.
\end{equation}
In terms of the potentials, taking into account that ${\kf}$ has the symmetries of the Riemann tensor (see the Appendix
), we obtain the modified Maxwell equations
\begin{equation}
\Box A_{\gamma}-\partial_{\gamma}(\partial^{\alpha}A_{\alpha})-2{\kf}_{\alpha\gamma\mu\nu}\partial^{\alpha}\partial^{\nu}A^{\mu}=0.\label{eq:smelageqmo}
\end{equation}
Proceeding in a fashion similar to those employed in quantizing the field potentials in the covariant theory~\cite{MandlAndShaw}, we introduce the SME Fermi Lagrangian
\begin{align}\begin{split}
\mathcal{L}=&\;-\frac{1}{2}(\partial_{\nu}A_{\mu})(\partial^{\nu}A^{\mu})-\frac{1}{4}{\kf}_{\kappa\lambda\mu\nu}F^{\kappa\lambda}F^{\mu\nu}\\
=&\;-\frac{1}{2}(\partial_{\nu}A_{\mu})(\partial^{\nu}A^{\mu})-{\kf}_{\kappa\lambda\mu\nu}(\partial^{\lambda}A^{\kappa})(\partial^{\nu}A^{\mu}),\label{eq:smefermilagrangian}
\end{split}\end{align}
which, like the fully Lorentz covariant Fermi Lagrangian used to quantize the covariant theory, has a nonzero momentum $\pi^{0}$ conjugate to $A^{0}$.
The equations of motion resulting from \eqref{eq:smefermilagrangian} are then
\begin{equation}
\Box A_{\gamma}-2{\kf}_{\alpha\gamma\mu\nu}(\partial^{\alpha}\partial^{\nu}A^{\mu})=0,\label{eq:smefermilageqmo}
\end{equation}
which are equivalent to \eqref{eq:smelageqmo}, provided that we enforce the Lorenz gauge condition 
\begin{equation}
{\partial^{\alpha}A_{\alpha}=0}.\label{eq:lorenzgauge}
\end{equation}
Separating the spatial and time-derivatives in the Lagrangian, we obtain
\begin{align}\begin{split}
\mathcal{L}=&\;-\frac{1}{2}\left[(\partial_{0}A_{\mu})(\partial^{0}A^{\mu})+(\partial_{p}A_{\mu})(\partial^{p}A^{\mu})\right]\\
&\;-{\kf}_{\kappa 0 \mu 0}(\partial^{0}A^{\kappa})(\partial^{0}A^{\mu})-{\kf}_{\kappa p \mu 0}(\partial^{p}A^{\kappa})(\partial^{0}A^{\mu})\\
&\;-{\kf}_{\kappa 0 \mu q}(\partial^{0}A^{\kappa})(\partial^{q}A^{\mu})-{\kf}_{\kappa p \mu q}(\partial^{p}A^{\kappa})(\partial^{q}A^{\mu})
\end{split}\end{align}
The full Lagrangian is obtained by integrating $\mathcal{L}$ over all space, so we may use the Parseval-Plancherel identity to obtain the reciprocal-space Lagrangian density
\begin{align}\begin{split}
\widetilde{\mathcal{L}}=&-\frac{1}{2}\left[(\partial_{0}\kAd{\mu}{k})(\partial^{0}\kAu{\mu}{k})^{*}+k_{p}k^{p}\kAd{\mu}{k}\kAu{\mu}{k}^{*}\right]\\
&\;-{\kf}_{\kappa 0 \mu 0}(\partial^{0}\kAu{\kappa}{k})(\partial^{0}\kAu{\mu}{k})^{*}\\
&\;+ik^{p}{\kf}_{\kappa p \mu 0}(\kAu{\kappa}{k})(\partial^{0}\kAu{\mu}{k})^{*}\\
&\;-ik^{q}{\kf}_{\kappa 0 \mu q}(\partial^{0}\kAu{\kappa}{k})(\kAu{\mu}{k})^{*}\\
&\;-k^{p}k^{q}{\kf}_{\kappa p \mu q}(\kAu{\kappa}{k})(\kAu{\mu}{k})^{*},
\end{split}\end{align}
from which the full Lagrangian may be recovered by integrating over all $\vec{k}$.  Because the potentials are real, we have
\begin{equation}
\kAu{\mu}{k}=\kAu{\mu}{-k}^{*},\label{eq:realpot}
\end{equation}
which permits us to write the full Lagrangian as an integral over only half of reciprocal space of the Lagrangian density $\widetilde{\mathcal{L}}_{R}$,
\begin{align}\begin{split}
\widetilde{\mathcal{L}}_{R}=&-\left[(\partial_{0}\kAd{\mu}{k})(\partial^{0}\kAu{\mu}{k})^{*}+k_{p}k^{p}\kAd{\mu}{k}\kAu{\mu}{k}^{*}\right]\\
&\;-{\kf}_{\kappa 0 \mu 0}(\partial^{0}\kAu{\kappa}{k})(\partial^{0}\kAu{\mu}{k})^{*}\\
&\;-{\kf}_{\kappa 0 \mu 0}(\partial^{0}\kAu{\kappa}{k})^{*}(\partial^{0}\kAu{\mu}{k})\\
&\;+ik^{p}{\kf}_{\kappa p \mu 0}(\kAu{\kappa}{k})(\partial^{0}\kAu{\mu}{k})^{*}\\
&\;-ik^{p}{\kf}_{\kappa p \mu 0}(\kAu{\kappa}{k})^{*}(\partial^{0}\kAu{\mu}{k})\\
&\;-ik^{q}{\kf}_{\kappa 0 \mu q}(\partial^{0}\kAu{\kappa}{k})(\kAu{\mu}{k})^{*}\\
&\;+ik^{q}{\kf}_{\kappa 0 \mu q}(\partial^{0}\kAu{\kappa}{k})^{*}(\kAu{\mu}{k})\\
&\;-k^{p}k^{q}{\kf}_{\kappa p \mu q}(\kAu{\kappa}{k})(\kAu{\mu}{k})^{*}\\
&\;-k^{p}k^{q}{\kf}_{\kappa p \mu q}(\kAu{\kappa}{k})^{*}(\kAu{\mu}{k}).\label{eq:LrLagrange}
\end{split}\end{align}
Taking $\kAu{\gamma}{k}$ as our coordinates, we find that the conjugate momenta are given by (using $\kpd{\gamma}{k}=\nolinebreak(1/c)\partial\widetilde{\mathcal{L}}_{R}/\partial(\partial^{0}\kAu{\gamma}{k}^{*})$):
\begin{align}\begin{split}
c\kpd{\gamma}{k}=&\;-(\partial_{0}\kAd{\gamma}{k})-2{\kf}_{ \gamma 0 \kappa 0 }(\partial^{0}\kAu{\kappa}{k})\\
&\;+2ik^{p}{\kf}_{\gamma 0 \kappa p}\kAu{\kappa}{k}.
\end{split}\end{align}
This can be solved to leading order in ${\kf}$ for $(\partial_{0}\kAd{\gamma}{k})$ as
\begin{align}\begin{split}
\partial_{0}\kAd{\gamma}{k}=&\;-c\kpd{\gamma}{k}+2c{\kf}_{ \gamma 0 \kappa 0 }\kpu{\kappa}{k}\\
&\;+2ik^{p}{\kf}_{\gamma 0 \kappa p}\kAu{\kappa}{k}+\mathcal{O}\left({\kf}^{2}\right).\label{eq:doAk}
\end{split}\end{align}
By substituting the leading order expansion \eqref{eq:doAk} for $(\partial_{0}\kAd{\gamma}{k})$ in \eqref{eq:LrLagrange}, we exchange the exact Lagrangian for one which is equivalent to first order in ${\kf}$ at the cost of adding additional unphysical terms at second order.  We seek a leading order expansion, and so shall ignore all second order and higher couplings.  This leads to the approximate Lagrangian density
\begin{multline}
\widetilde{\mathcal{L}}_{R}=\left[c^{2}(2{\kf}_{\kappa 0 \mu 0}-g_{\kappa\mu})\kpu{\kappa}{k}\kpu{\mu}{k}^{*}-\right.\\
\left.\left(g_{\kappa\mu}g_{pq}+2{\kf}_{\kappa p \mu q}\right)k^{p}k^{q}(\kAu{\kappa}{k})(\kAu{\mu}{k})^{*}\right],\label{eq:leadingorderLag}
\end{multline}
where $g_{\mu\nu}$ is the Minkowski metric: $g_{\mu\nu}=\text{diag}\left(1,-1,-1,-1\right)$.
\section{The Hamiltonian\label{sec:Hamiltonian}}
The Hamiltonian density is given by
\begin{equation}
\widetilde{\mathcal{H}}_{R}=c(\kpu{\gamma}{k})(\partial_{0}\kAd{\gamma}{k})^{*}+c(\partial_{0}\kAd{\gamma}{k})(\kpu{\gamma}{k})^{*}-\widetilde{\mathcal{L}}_{R},
\end{equation}
and so using \eqref{eq:doAk} and \eqref{eq:leadingorderLag}, $\widetilde{\mathcal{H}}_{R}$ becomes
\begin{multline}
\widetilde{\mathcal{H}}_{R}=\left((g_{\kappa\mu}g_{pq}+2{\kf}_{\kappa p \mu q})k^{p}k^{q}(\kAu{\kappa}{k})(\kAu{\mu}{k})^{*}\right.\\
\left.-c^{2}(g_{\kappa\mu}-2{\kf}_{\kappa 0 \mu 0})\kpu{\kappa}{k}\kpu{\mu}{k}^{*}\right)\\
\;-2ick^{p}{\kf}_{\gamma 0 \kappa p}\left[(\kpu{\gamma}{k})(\kAu{\kappa}{k})^{*}-(\kAu{\kappa}{k})(\kpu{\gamma}{k})^{*}\right].\label{eq:hamiltonianR}
\end{multline}
Since this theory is a perturbation of the fully Lorentz covariant theory, we expect the normal modes that result to be perturbations of the fully covariant normal modes.  These standard normal modes can be written in terms of $\kAu{\mu}{k}$ and $\kpu{\mu}{k}$, so that
\begin{eqnarray}
\alpha^{\mu}(\vec{k})&=&\sqrt{\frac{c^{2}}{2\hbar \omega_{k}}}\left[\frac{\omega_{k}}{c^{2}}\kAu{\mu}{k}+i\kpu{\mu}{k}\right]\\
\alpha^{\mu}(\vec{k})^{*}&=&
\sqrt{\frac{c^{2}}{2\hbar\omega_{k}}}\left[\frac{\omega_{k}}{c^{2}}\kAu{\mu}{-k}-i\kpu{\mu}{-k}\right]\\
\alpha^{\mu}(-\vec{k})&=&\sqrt{\frac{c^{2}}{2\hbar \omega_{k}}}\left[\frac{\omega_{k}}{c^{2}}\kAu{\mu}{-k}+i\kpu{\mu}{-k}\right]\\
\alpha^{\mu}(-\vec{k})^{*}&=&
\sqrt{\frac{c^{2}}{2\hbar \omega_{k}}}\left[\frac{\omega_{k}}{c^{2}}\kAu{\mu}{k}-i\kpu{\mu}{k}\right],
\end{eqnarray}
where we have made use of the reality of the potentials and their conjugate momenta \eqref{eq:realpot}.  Note that insofar as choosing a set of variables to write the Hamiltonian in terms of, we are free to make use of any linear combination of $\kAu{\mu}{k}$ and $\kpu{\mu}{k}$ that yield an acceptable commutator.  We have chosen $\omega_{k}=|\vec{k}|c$, as is usual for the fully covariant theory.  Since our choice of $\omega_{k}$ does not necessarily satisfy the Lorentz-violating dispersion relation, there will be terms coupling the forward propagating modes to those propagating backwards in the Hamiltonian.  At the end of this derivation, these and other like terms will ultimately be eliminated by a transformation of the mode operators which diagonalizes $\widetilde{\mathcal{H}}_{R}$, and which can be interpreted in part as changing $\omega_{k}$ to satisfy the appropriate dispersion relation.  Proceeding using this set of (approximately) normal modes, we then find that
\begin{eqnarray}
\kAu{\mu}{k}&=&\sqrt{\frac{\hbar c^{2}}{2\omega_{k}}}\left(\alpha^{\mu}(\vec{k})+\alpha^{\mu}(-\vec{k})^{*}\right)\\
\kpu{\mu}{k}&=&-i\sqrt{\frac{\hbar \omega_{k}}{2c^{2}}}\left(\alpha^{\mu}(\vec{k})-\alpha^{\mu}(-\vec{k})^{*}\right).
\end{eqnarray}

We can quantize this theory by identifying $\kAu{\mu}{k}$ and $\kpu{\nu}{k}$ as operators with the canonical commutation relation
\begin{eqnarray}
\left[\kAu{\mu}{k},\overline{\pi^{\nu}}(\vec{k'})\right]=i\hbar g^{\mu\nu}\delta(\vec{k}-\vec{k}'),\label{eq:kAcommPi}
\end{eqnarray}
where $\overline{A}$ represents the adjoint of an operator $A$.  We use this peculiar form so as to be consistent with the notation of \cite{CohenTannoudjiPhotonsAndAtoms}, and to distinguish the properties of the adjoint in the canonically quantized metric from those of the adjoint in the ``physical'' metric used to define a basis in Hilbert space, as discussed in more detail in section \ref{sec:whyindef}.  The (approximately) normal modes $\alpha(\vec{k})$ now become operators $a(\vec{k})$, whose non-vanishing commutators are, from \eqref{eq:kAcommPi}
\begin{equation}
\left[a_{r}(\vec{k}),\ab{s}{k'}\right]=\zeta_{r}\delta_{rs}\delta(\vec{k}-\vec{k}'),\label{eq:arasd}
\end{equation}
where $\zeta_{r}=\{-1,1,1,1\}$ for $r=\{0,1,2,3\}$~\cite{note2}.  In what follows, it will be useful to distinguish between the scalar, transverse, and longitudinal modes associated with a given $\vec{k}$.  Thus we take the $\{\ao{0}{k},\ab{0}{k}\}$ to act on the scalar modes, $\{\ao{3}{k},\ab{3}{k}\}$ to act on longitudinal modes, and the $\{\ao{1}{k},\ab{1}{k}\}$ and $\{\ao{2}{k},\ab{2}{k}\}$ operators to act on the transverse modes for fields propagating parallel to $\vec{k}$.  We can then write $\kA$ and $\pi$ as
\begin{eqnarray}
\kA^{\mu}(\vec{k})&=&\sqrt{\frac{\hbar c^{2}}{2\omega_{{k}}}}\sum_{r}\epsilon_{r}^{\mu}(\vec{k})\left(\ao{r}{k}+\ab{r}{-k}\right)\label{eq:kAaad}\\
\pi^{\nu}(\vec{k})&=&-i\sqrt{\frac{\hbar\omega_{{k}}}{2c^{2}}}\sum_{s}\epsilon_{s}^{\nu}(\vec{k})\left( \ao{s}{k}-\ab{s}{-k}\right)\label{eq:kpiaad}.
\end{eqnarray}
The newly introduced $\epsilon_{s}^{\nu}(\vec{k})$ tensor is responsible for keeping track of which time-spatial components of $A^{\mu}$ are excited by the mode operators.  Following \cite{MandlAndShaw}, and as defined in part \ref{sec:smephoton}, $\epsilon_{0}^{\nu}(\vec{k})=(-1,0,0,0)$, while the spatial components $\epv{j}{k}$ form a set of mutually orthogonal polarization vectors for each $\vec{k}$, defined in Eq.~\eqref{eq:polarizationdefs}. 
With these definitions, \eqref{eq:arasd} is easily shown to be consistent with \eqref{eq:kAcommPi}.  Note that at this point, we can immediately infer that the form of the fields' conserved momentum operator is unchanged from its form in the fully covariant theory, since the conserved momentum density is given by
\begin{equation}
\mathcal{P}^{j}(\vec{k})=\pi_{\gamma}(\vec{k})\left(-i\vec{k}^{j}\kAu{\gamma}{k}\right)^{*}+\pi_{\gamma}(\vec{k})^{*}\left(i\vec{k}^{j}\kAu{\gamma}{k}\right),
\end{equation}
which does not depend upon $\kf$.  Substituting \eqref{eq:kAaad} and \eqref{eq:kpiaad} into \eqref{eq:hamiltonianR}, we find 
\begin{widetext}
\begin{align}\begin{split}
\widetilde{\mathcal{H}}_{R}=&\;\hbar\omega_{k}\left(-\epsilon_{r,\mu}(\vec{k})\epsilon_{s}^{\mu}(\vec{k})\right)\left(\ao{r}{k}\ab{s}{k}+\ab{r}{-k}\ao{s}{-k}\right)\\
&\;+\hbar\omega_{k}\left(\epsilon_{r}^{\kappa}(\vec{k})\epsilon_{s}^{\mu}(\vec{k})\right)\left({\kf}_{\kappa p \mu q}\hat{k}^{p}\hat{k}^{q} +{\kf}_{\kappa 0 \mu 0}-{\kf}_{\mu 0 \kappa p}\hat{k}^{p}\right)\left[\ao{r}{k}\ab{s}{k}\right]\\
&\;+\hbar\omega_{k}\left(\epsilon_{r}^{\kappa}(\vec{k})\epsilon_{s}^{\mu}(\vec{k})\right)\left({\kf}_{\kappa p \mu q}\hat{k}^{p}\hat{k}^{q} +{\kf}_{\kappa 0 \mu 0}+{\kf}_{\mu 0 \kappa p}\hat{k}^{p}\right)\left[\ab{r}{-k}\ao{s}{-k}\right]\\
&\;+\hbar\omega_{k}\left(\epsilon_{r}^{\kappa}(\vec{k})\epsilon_{s}^{\mu}(\vec{k})\right)\left({\kf}_{\kappa p \mu q}\hat{k}^{p}\hat{k}^{q} -{\kf}_{\kappa 0 \mu 0}+{\kf}_{\mu 0 \kappa p}\hat{k}^{p}\right)\left[\ao{r}{k}\ao{s}{-k}\right]\\
&\;+\hbar\omega_{k}\left(\epsilon_{r}^{\kappa}(\vec{k})\epsilon_{s}^{\mu}(\vec{k})\right)\left({\kf}_{\kappa p \mu q}\hat{k}^{p}\hat{k}^{q} -{\kf}_{\kappa 0 \mu 0}-{\kf}_{\mu 0 \kappa p}\hat{k}^{p}\right)\left[\ab{r}{-k}\ab{s}{k}\right]\\
&\;-\hbar \omega_{k}\hat{k}^{p}{\kf}_{\mu 0 \kappa p}\left(\epsilon_{r}^{\kappa}(\vec{k})\epsilon_{s}^{\mu}(\vec{k})\right)\left[\ao{s}{k}\ab{r}{k}-\ab{s}{-k}\ao{r}{-k}\right]\\
&\;-\hbar \omega_{k}\hat{k}^{p}{\kf}_{\mu 0 \kappa p}\left(\epsilon_{r}^{\kappa}(\vec{k})\epsilon_{s}^{\mu}(\vec{k})\right)\left[\ao{s}{k}\ao{r}{-k}-\ab{s}{-k}\ab{r}{k}\right]\label{eq:hamiltonianRII}.
\end{split}\end{align}
\end{widetext}
The first line of the above expression for $\widetilde{\mathcal{H}}_{R}$ is that of the covariant free-field, and the terms that follow represent the Lorentz-violating perturbation.
Making use of the identity \eqref{eq:kfVectorIdent} in in the Appendix, 
we find that
\begin{multline}
(k_{F})_{\kappa 0 \mu 0}\epsilon_{r}^{\kappa}(\vec{k})\epsilon_{s}^{\mu}(\vec{k})=\\
-\frac{1}{2}\left[\vec{\epsilon}_{r}(\vec{k})\cdot\kemktr\cdot\vec{\epsilon}_{s}(\vec{k})\right],
\end{multline}
\begin{multline}
\epsilon_{r}^{\kappa}(\vec{k})\epsilon_{s}^{\mu}(\vec{k})(k_{F})_{\kappa p \mu q}\hat{k}^{p}\hat{k}^{q}=\\
-\frac{1}{2}\Bigg\{\epsilon_{r}^{0}(\vec{k})\epsilon_{s}^{0}(\vec{k})\left[\epv{3}{k}\cdot\kemktr\cdot\epv{3}{k}\right]\\
\:\:\:\:\:\:+\epsilon_{s3m}\epsilon_{r}^{0}(\vec{k})\left[\epv{3}{k}\cdot\kop\cdot\epv{m}{k}\right]\\
\:\:\:\:\:\:+\epsilon_{r3m}\epsilon_{s}^{0}(\vec{k})\left[\epv{3}{k}\cdot\kop\cdot\epv{m}{k}\right]\\
\:\:\:\:\:\:+\epsilon_{r3n}\epsilon_{s3m}\left[\epv{n}{k}\cdot\kemktr\cdot\epv{m}{k}\right]\Bigg\},
\end{multline}
and
\begin{multline}
\epsilon_{r}^{\kappa}(\vec{k})\epsilon_{s}^{\mu}(\vec{k})(k_{F})_{\mu 0 \kappa p}\hat{k}^{p}=\\
\;\frac{1}{2}\bigg(\epsilon_{r}^{0}(\vec{k})\left[\epv{s}{k}\cdot\kemktr\cdot\epv{3}{k}\right]\\
\:\:+\epsilon_{r3m}\left[\epv{s}{k}\cdot\kop\cdot\epv{m}{k}\right]\bigg).
\end{multline}
By substituting the above three expressions into the Hamiltonian \eqref{eq:hamiltonianRII}, and taking full advantage of the symmetry of the $\kem$ matrix and the antisymmetry of $\kop$; which is such that the scalar product with two vectors $\vec{v}_{1}$ and $\vec{v}_{2}$ obey
\begin{eqnarray}
\vec{v}_{1}\cdot\kem\cdot\vec{v}_{2}&=&\vec{v}_{2}\cdot\kem\cdot\vec{v}_{1}\\
\vec{v}_{1}\cdot\kop\cdot\vec{v}_{2}&=&-\vec{v}_{2}\cdot\kop\cdot\vec{v}_{1},
\end{eqnarray}
we can write the Hamiltonian density in five parts.  
\begin{equation}
\widetilde{\mathcal{H}}_{R}=\wtcal{H}_{T}+\wtcal{H}_{\pm,T}+\wtcal{H}_{LS}+\wtcal{H}_{+,T,LS}+\wtcal{H}_{-,T,LS},
\end{equation}
where $\wtcal{H}_{T}$ includes products of transverse mode operators with the same wavevector $\vec{k}$, $\wtcal{H}_{\pm,T}$ contains products of the transverse mode operators with opposing wavevectors $-\vec{k}$, $\wtcal{H}_{LS}$ includes terms involving only the longitudinal and scalar modes, and the couplings between the ``positive'' and ``negative'' transverse modes with the longitudinal and scalar degrees of freedom are expressed in $\wtcal{H}_{+,T,LS}$ and $\wtcal{H}_{-,T,LS}$.  To simplify the expression for $\wtcal{H}_{T}$, we write the fractional shift in the speed of light moving parallel to $\vec{k}$ due to the Lorentz-violating terms as
\begin{multline}
\delta(\vec{k})=\epvkopepv{1}{2}{k}\\
-\frac{1}{2}\sum_{r=1}^{2}\epvkemktrepv{r}{r}{k}.
\end{multline}
Recalling that the Hamiltonian density $\wtcal{H}_{R}$ is only summed over half of reciprocal space, we obtain
\begin{align}\begin{split}
\wtcal{H}_{T}=&\;\hbar\omega_{k}\left[1+\delta(\vec{k})\right]\left(\ao{1}{k}\ab{1}{k}+\ao{2}{k}\ab{2}{k}\right)\\
&+\hbar\omega_{k}\left[1+\delta(-\vec{k})\right]\left(\ab{1}{-k}\ao{1}{-k}+\ab{2}{-k}\ao{2}{-k}\right).\label{eq:Htransverse}
\end{split}\end{align}
This shows that the leading order shift to the energy of photons with wavevector $\vec{k}$ is consistent with the dispersion relation derived from the Lagrangian \cite{Colladay:1998,Kostelecky:2002}.  The remaining $\wtcal{H}_{\pm,T}$, $\wtcal{H}_{+,T,LS}$, and $\wtcal{H}_{-,T,LS}$ terms, as well as the cross couplings between scalar and longitudinal modes in $\wtcal{H}_{LS}$, can be attributed to the differences between the normal modes of the covariant theory and those of the Lorentz-violating model, and are given by.
\begin{widetext}
\begin{align}\begin{split}
\wtcal{H}_{\pm,T}=&\;\frac{\hbar\omega_{k}}{2}\bigg\{\left(\epvkemktrepv{1}{1}{k}-\epvkemktrepv{2}{2}{k}        \right)\times\\
&\:\:\:\:\:\:\:\:\:\:\:\:\:\:\:\:\:\left(\ao{1}{k}\ao{1}{-k}+\ab{1}{-k}\ab{1}{k}-\ao{2}{k}\ao{2}{-k}-\ab{2}{-k}\ab{2}{k}\right)\\
&\phantom{\frac{\hbar\omega_{k}}{2}\bigg\{}+\left(2\epvkemktrepv{1}{2}{k}        \right)\left(\ao{1}{k}\ao{2}{-k}+\ab{2}{-k}\ab{1}{k}\right)\\
&\phantom{\frac{\hbar\omega_{k}}{2}\bigg\{}+\left(2\epvkemktrepv{1}{2}{k}        \right)\left(\ao{2}{k}\ao{1}{-k}+\ab{1}{-k}\ab{2}{k}\right)\bigg\}.
\end{split}\end{align}

\begin{align}\begin{split}
\wtcal{H}_{LS}=&\;\hbar\omega_{k}\left(\ao{3}{k}\ab{3}{k}+\ab{3}{-k}\ao{3}{-k}\right)-\left(\ao{0}{k}\ab{0}{k}+\ab{0}{-k}\ao{0}{-k}\right)\\
&-\frac{\hbar\omega_{k}}{2}\epvkemktrepv{3}{3}{k}\bigg\{\ao{3}{k}\ab{3}{k}+\ab{3}{-k}\ao{3}{-k}\\
&\:\:\:\:\:\:\:\:\:\:\:\:\:\:\:+\ao{0}{k}\ab{0}{k}+\ab{0}{-k}\ao{0}{-k}+\ao{0}{k}\ao{0}{-k}+\ab{0}{-k}\ab{0}{k}\\
&\:\:\:\:\:\:\:\:\:\:\:\:\:\:\:+\ao{3}{k}\ao{0}{-k}+\ab{0}{-k}\ab{3}{k}-\ao{0}{k}\ao{3}{-k}-\ab{3}{-k}\ab{0}{k}\\
&\:\:\:\:\:\:\:\:\:\:\:\:\:\:\:-\ao{3}{k}\ao{3}{-k}-\ab{3}{-k}\ab{3}{k}+\ao{0}{k}\ab{3}{k}+\ao{3}{k}\ab{0}{k}\\
&\:\:\:\:\:\:\:\:\:\:\:\:\:\:\:-\ab{0}{-k}\ao{3}{-k}-\ab{3}{-k}\ao{0}{-k}\bigg\}
\end{split}\end{align}
\begin{align}\begin{split}
\wtcal{H}_{+,T,LS}=&\;\frac{-\hbar\omega_{k}}{2}\bigg\{\left(\epvkemktrepv{1}{3}{k}-\epvkopepv{3}{2}{k}\right)\times\\
&\:\:\:\:\:\:\:\:\:\:\:\:\left(\ao{0}{k}\ab{1}{k}+\ao{1}{k}\ab{0}{k}+\ao{3}{k}\ab{1}{k}+\ao{1}{k}\ab{3}{k}\right.\\
&\:\:\:\:\:\:\:\:\:\:\:\:\:\:\:\:\:\left.+\ao{1}{k}\ao{0}{-k}+\ab{0}{-k}\ab{1}{k}-\ao{1}{k}\ao{3}{-k}-\ab{3}{-k}\ab{1}{k}\right)\\
&\phantom{-\frac{\hbar\omega_{k}}{2}\bigg\{}+\left(\epvkemktrepv{2}{3}{k}+\epvkopepv{3}{1}{k}\right)\times\\
&\:\:\:\:\:\:\:\:\:\:\:\:\left(\ao{0}{k}\ab{2}{k}+\ao{2}{k}\ab{0}{k}+\ao{3}{k}\ab{2}{k}+\ao{2}{k}\ab{3}{k}\right.\\
&\:\:\:\:\:\:\:\:\:\:\:\:\:\:\:\:\:\left.+\ao{2}{k}\ao{0}{-k}+\ab{0}{-k}\ab{2}{k}-\ao{2}{k}\ao{3}{-k}-\ab{3}{-k}\ab{2}{k}\right)\bigg\}
\end{split}\end{align}
\begin{align}\begin{split}
\wtcal{H}_{-,T,LS}=&\;\frac{\hbar\omega_{k}}{2}\bigg\{\left(\epvkemktrepv{1}{3}{k}+\epvkopepv{3}{2}{k}\right)\times\\
&\:\:\:\:\:\:\:\:\:\:\:\:\left(\ab{0}{-k}\ao{1}{-k}+\ab{1}{-k}\ao{0}{-k}-\ab{3}{-k}\ao{1}{-k}-\ab{1}{-k}\ao{3}{-k}\right.\\
&\:\:\:\:\:\:\:\:\:\:\:\:\:\:\:\:\:\left.+\ao{0}{k}\ao{1}{-k}+\ab{1}{-k}\ab{0}{k}+\ao{3}{k}\ao{1}{-k}+\ab{1}{-k}\ab{3}{k}\right)\\
&\phantom{-\frac{\hbar\omega_{k}}{2}\bigg\{}+\left(\epvkemktrepv{2}{3}{k}-\epvkopepv{3}{1}{k}\right)\times\\
&\:\:\:\:\:\:\:\:\:\:\:\:\left(\ab{0}{-k}\ao{2}{-k}+\ab{2}{-k}\ao{0}{-k}-\ab{3}{-k}\ao{2}{-k}-\ab{2}{-k}\ao{3}{-k}\right.\\
&\:\:\:\:\:\:\:\:\:\:\:\:\:\:\:\:\:\left.+\ao{0}{k}\ao{2}{-k}+\ab{2}{-k}\ab{0}{k}+\ao{3}{k}\ao{2}{-k}+\ab{2}{-k}\ab{3}{k}\right)\bigg\}.
\end{split}\end{align}
\end{widetext}
The Hamiltonian can be further simplified by expressing the scalar and longitudinal operators in terms of 
\begin{eqnarray}
\ao{d}{k}&=&\frac{i}{\sqrt{2}}\left(\ao{3}{k}-\ao{0}{k}\right)\\
\ao{g}{k}&=&\frac{1}{\sqrt{2}}\left(\ao{3}{k}+\ao{0}{k}\right),
\end{eqnarray}
so that
\begin{widetext}
\begin{align}\begin{split}
\wtcal{H}_{LS}=&\;\;i\hbar\omega_{k}\left(\ao{d}{k}\ab{g}{k}-\ao{g}{k}\ab{d}{k}+\ab{g}{-k}\ao{d}{-k}-\ab{d}{-k}\ao{g}{-k}\right)\\
&-\hbar\omega_{k}\epvkemktrepv{3}{3}{k}\left(\ao{g}{k}\ab{g}{k}+\ab{d}{-k}\ao{d}{-k}\right)\\
&-i\hbar\omega_{k}\epvkemktrepv{3}{3}{k}\left(\ao{g}{k}\ao{d}{-k}-\ab{d}{-k}\ab{g}{k}\right)
\label{eq:hamLS}
\end{split}\end{align}
\begin{align}\begin{split}
\wtcal{H}_{+,T,LS}=&\;\frac{-\hbar\omega_{k}}{\sqrt{2}}\bigg\{\left(\epvkemktrepv{1}{3}{k}-\epvkopepv{3}{2}{k}\right)\times\\
&\:\:\:\:\:\:\:\:\:\:\:\:\:\:\:\:\:\:\:\:\:\:\left(\ao{1}{k}\left[\ab{g}{k}+i\ao{d}{-k}\right]+\left[\ao{g}{k}-i\ab{d}{-k}\right]\ab{1}{k}\right)\\
&\phantom{-\frac{\hbar\omega_{k}}{2}\bigg\{}+\left(\epvkemktrepv{2}{3}{k}+\epvkopepv{3}{1}{k}\right)\times\\
&\:\:\:\:\:\:\:\:\:\:\:\:\:\:\:\:\:\:\:\:\:\:\left(\ao{2}{k}\left[\ab{g}{k}+i\ao{d}{-k}\right]+\left[\ao{g}{k}-i\ab{d}{-k}\right]\ab{2}{k}\right)\bigg\},\label{eq:hamPTLS}
\end{split}\end{align}
\begin{align}\begin{split}
\wtcal{H}_{-,T,LS}=&\;\frac{\hbar\omega_{k}}{\sqrt{2}}\bigg\{\left(\epvkemktrepv{1}{3}{k}+\epvkopepv{3}{2}{k}\right)\times\\
&\:\:\:\:\:\:\:\:\:\:\:\:\:\:\:\:\:\:\:\:\:\:\left(\left[\ao{g}{k}-i\ab{d}{-k}\right]\ao{1}{-k}+\ab{1}{-k}\left[\ab{g}{k}+i\ao{d}{-k}\right]\right)\\
&\phantom{-\frac{\hbar\omega_{k}}{2}\bigg\{}+\left(\epvkemktrepv{2}{3}{k}-\epvkopepv{3}{1}{k}\right)\times\\
&\:\:\:\:\:\:\:\:\:\:\:\:\:\:\:\:\:\:\:\:\:\:\left(\left[\ao{g}{k}-i\ab{d}{-k}\right]\ao{2}{-k}+\ab{2}{-k}\left[\ab{g}{k}+i\ao{d}{-k}\right]\right)\bigg\}.\label{eq:hamMTLS}
\end{split}\end{align}
To leading order in $\tilde{\kappa}$, the interactions between the transverse modes contained in $\wtcal{H}_{\pm,T}$ can be eliminated by performing the unitary transformation 
\begin{equation}
e^{\Xi_{1}+\Xi_{2}}\,\wtcal{H}_{R}\,e^{-\Xi_{1}-\Xi_{2}},\label{eq:diagtransverse}
\end{equation}
where
\begin{align}\begin{split}
\Xi_{1}=&\;\sum_{\vec{k}}\frac{1}{4}\left(\epvkemktrepv{1}{1}{k}\right.\\
&\left.-\epvkemktrepv{2}{2}{k}\right)\times\\
&
\left(\ab{1}{k}\ab{1}{-k}-\ao{1}{-k}\ao{1}{k}-\ab{2}{k}\ab{2}{-k}+\ao{2}{-k}\ao{2}{k}\right)\\
\Xi_{2}=&\;\sum_{\vec{k}}\frac{1}{2}\epvkemktrepv{1}{2}{k}\times\\
&\:\:\:\:\:\:\:\:\:\:\:\:\:\:\:\:\:\:\:\:\:\:\:\:\left(\ab{1}{k}\ab{2}{-k}-\ao{1}{k}\ao{2}{-k}+\ab{2}{k}\ab{1}{-k}-\ao{2}{k}\ao{1}{-k}\right).
\end{split}\end{align}
\end{widetext}
Thus we may write the free field Hamiltonian in terms of \eqref{eq:Htransverse}, \eqref{eq:hamLS}, \eqref{eq:hamPTLS}, and \eqref{eq:hamMTLS} as
\begin{equation}
e^{\Xi_{1}+\Xi_{2}}\,\wtcal{H}_{R}\,e^{-\Xi_{1}-\Xi_{2}}=\wtcal{H}_{T}+\wtcal{H}_{LS}+\wtcal{H}_{+,T,LS}+\wtcal{H}_{-,T,LS}.\label{eq:DHam}
\end{equation}

As will be demonstrated in part \ref{subsec:LVHIM}, the remaining $\wtcal{H}_{LS}$ and $\wtcal{H}_{\pm,T,LS}$ terms do not contribute to physical observables, and do not affect the evolution of the free fields at leading order.  Thus the similarity transform \eqref{eq:diagtransverse} has effectively diagonalized the free-field Hamiltonian.  We note that at second order in $\tilde{\kappa}$, the $\wtcal{H}_{\pm,T,LS}$ terms can generate vacuum birefringence via an intermediate coupling to the scalar and longitudinal modes (\ie $d$- and $g$-modes).  This is qualitatively consistent with the solution to the Lagrangian equations of motion \eqref{eq:smefermilageqmo} taken to second order in $\kem$, $\kop$, and $\ktr$, although a rigorous treatment would require the inclusion of numerous second order terms (all of which are suppressed by at least a factor of $10^{12}$ relative to the leading order effects) which were discarded in the course of this derivation.  The detailed forms of $\wtcal{H}_{LS}$ and $\wtcal{H}_{\pm,T,LS}$ are of great importance in any fully quantum treatment of electro- and magneto-statics in the photon sector of the SME.

\renewcommand{\ao}[1]{\ensuremath{a_{#1}}}
\renewcommand{\ab}[1]{\bar{a}_{#1}}
\renewcommand{\ad}[1]{\ensuremath{a^{\dagger}_{#1}}}
\renewcommand{\abd}[1]{\bar{a}^{\dagger}_{#1}}

\section{The Indefinite Metric\label{sec:whyindef}}
While the Hamiltonian \eqref{eq:DHam} is self-adjoint in the sense that $\wtcal{H}_{R}=\overline{\wtcal{H}_{R}}$, this fact alone does not establish that eigenstates of $\wtcal{H}_{R}$ will satisfy the Lorenz condition, and thus represent solutions to the modified Maxwell equations.  In contrast to the fully covariant theory, \eqref{eq:DHam} includes a variety of terms coupling the physically permitted transverse modes to the unphysical scalar and longitudinal modes.  Here, we demonstrate that these terms do not couple states that are consistent with Maxwells equations to states that are not; and that \eqref{eq:DHam} is the operator of a well defined observable which can act as the generator of translations in time.  To do this, we follow the usual process by which the potentials of the fully covariant theory are quantized, and choose to define a basis for the quantized fields' Hilbert space in a metric other than the one induced by \eqref{eq:kAcommPi}.  

We first review the properties of the inner product, or metric, that covariant quantization imposes on the Hilbert space, and reprise the procedure by which the metric is redefined to permit the construction of a basis for the Hilbert space comprised of states with non-negative (if not strictly positive definite) norm.  For the fully covariant theory, this process is sufficient to completely isolate a subspace $S$ of states satisfying the Lorenz condition and that have positive norm from those that do not.  In the Lorentz-violating theory, however, the $\wtcal{H}_{\pm,T,LS}$ terms do not leave the subspace $S$ invariant.  Fortunately, as we will show in part \ref{subsec:LVHIM}, the Lorentz-violating theory leaves the larger subspace $S_{LV}$ of the states consistent with the modified Maxwell equations $S_{LV}\supset S$ invariant.  Although the metric on states in $S_{LV}$ is not strictly positive, it is non-negative.  We demonstrate that every $\ket{\psi}\in S_{LV}$ is a solution of the modified Maxwell equations \eqref{eq:smelageqmo}.  In so doing, we demonstrate that the form of the Lorenz condition used in the course of covariant quantization of the fully covariant theory is stronger than is strictly required, and develop a minimal ``weak'' Lorenz condition to define $S_{LV}$.  Finally, we show that to leading order in $\tilde{\kappa}$, states in $S_{LV}$ outside of $S$ can be ignored, and the metric can again be treated as if it were strictly positive.

\subsection{Origins of the Indefinite Metric\label{subsec:whyindef}}
In the process of covariant quantization, we made two fateful decisions.  First, we chose to quantize the \emph{potentials} $A^{\mu}$ and their conjugate momenta, rather than use the physically observable electric and magnetic fields.  This choice makes the interaction of the quantized field with Dirac fermions particularly straightforward, but inserts an additional unphysical degree of freedom into our system, associated with gauge invariance.  Next, in order to obtain a fully covariant commutation relation between the coordinate potentials and their conjugate momenta, we had to use a variant of the Fermi Lagrangian to induce a nonvanishing momentum for the time-component of the potential, inserting another degree of freedom.  This means that where we once had a system that admitted only transverse solutions of the free-field wave equation, we now have a representation of that system for which, in the absence of the appropriate constraints, scalar and longitudinal modes are permitted~\cite{note3}.  These unphysical degrees of freedom cause the Hilbert space of the quantized fields to include wavefunctions that are not solutions of \eqref{eq:smelageqmo}.  This problem can be addressed in more detail once we have constructed a suitable basis in part \ref{subsec:HilbertConStruct}.  Specifying  that basis in terms of the normal mode operators defined in \eqref{eq:kAaad} and \eqref{eq:kpiaad} is complicated by the covariant commutation relation between the potentials and their conjugate momenta: \begin{equation}
\left[A^{\mu}(\vec{r},t),\pi^{\nu}(\vec{r}\,',t')\right]=i\hbar g^{\mu\nu}\delta(t-t')\delta(\vec{r}-\vec{r}\,').
\end{equation}
As stated in \eqref{eq:arasd}, this gives rise to the equal time commutation relation between the normal modes in reciprocal space
\begin{equation}
\left[a_{r}(\vec{k}),\overline{a}_{s}(\vec{k}')\right]=\zeta_{r}\delta_{rs}\delta(\vec{k}-\vec{k}'),\label{eq:aadcommutator}
\end{equation}
with $\zeta_{r}=\{-1,1,1,1\}$ for $r=\{0,1,2,3\}$.  Because $[a_{0}(\vec{k}),a_{0}^{\dagger}(\vec{k})]=-1$, respectively identifying $\ad{0}$ and $\ao{0}$ as creation and annihilation operators leads to states with negative norm.  If the vacuum is normalized such that $\langle 0|0\rangle=1$, then one such negative norm state is that  with a single scalar-mode photon
\begin{equation}
\langle 1_{0}|1_{0}\rangle = \langle 0|a_{0}a_{0}^{\dagger}|0\rangle=-\langle 0|0\rangle+\langle0 |a_{0}^{\dagger}a_{0}|0\rangle=-1.\label{eq:indefnegnorm}
\end{equation}
This is a direct consequence of quantizing the potentials of the Fermi Lagrangian, which has led to a Hilbert space with an \emph{indefinite} (rather than strictly positive) inner product, or metric.  



\subsection{Properties of the Indefinite Metric}

Paralleling the discussion in \cite{CohenTannoudjiPhotonsAndAtoms}, we can define a new metric with respect to an existing Hilbert space (whose elements are denoted as $\ket{\psi}$) in terms of an operator $M$, hermitian on all $\ket{\psi}$, such that $M=M^{\dagger}=M^{-1}$.  Using this \emph{metric operator} $M$, we can then define a new metric on the Hilbert space in terms of $\ket{\psi}$ and the original metric by
\begin{equation}
\ibraket{\psi}{\phi} = \langle \psi| M|\phi\rangle,
\end{equation}
where $\iket{}$ and $\ibra{}$ are isomorphic to the physical states according to
\begin{equation}
\iket{\psi} = \ket{\psi}\text{, and }\ibra{\psi}=\bra{\psi}M.
\end{equation}
This implies that
\begin{equation}
\ibraket{\psi}{\phi} = \bra{\psi}M\ket{\phi} = \left(\bra{\phi}M^{\dagger}\ket{\psi}\right)^{*}=\left(\ibraket{\phi}{\psi}\right)^{*}.
\end{equation}
As was the case in the original metric, the product $\ibraket{\psi}{\phi}$ is linear in $\iket{\phi}$ and antilinear in $\ibra{\psi}$.  Even though we may initially choose $\bra{\psi}\psi\rangle$ to be positive definite, $\ibraket{\psi}{\psi}$ need not be, since
\begin{multline}
\ibraket{\psi}{\psi}=\bra{\psi}M\ket{\psi}
=\bra{\psi}\left(\sum_{j}m_{j}\ket{m_{j}}\bra{m_{j}}\right)\ket{\phi}\\
=\sum_{j}m_{j}|\bra{\psi}m_{j}\rangle|^{2},
\end{multline}
and the eigenvalues $m_{j}$ of $M$ can be $\pm1$, leading to the possibility of states with vanishing or negative norm.  If the original metric is positive definite, then metrics derived from that metric by a metric operator $M$ with one or more negative eigenvalues are termed \emph{indefinite}.  The freedom to choose $M$ permits us to define a new adjoint $\bar{A}$ such that
\begin{equation}
\ibra{\psi}A\iket{\phi}\; = \left(\ibra{\phi}\bar{A}\iket{\psi}\right)^{*}\label{eq:baradjointdef}
\end{equation}
is satisfied.  The new adjoint can be related to the old adjoint via
\begin{multline}
\ibra{\psi}A\iket{\phi}=\bra{\psi}MA\ket{\phi}=\bra{\psi}\bar{A}^{\dagger}M^{\dagger}\ket{\phi}\\
=\left(\bra{\phi}M\bar{A}\ket{\psi}\right)^{*}=\left(\ibra{\phi}\bar{A}\iket{\psi}\right)^{*},
\end{multline}
which implies $A^{\dagger}M^{\dagger}=M\bar{A}$.  Since $M=M^{\dagger}$ and $M^{2}=I$, we have that the new adjoint is given by
\begin{equation}
\bar{A}=MA^{\dagger}M.\label{eq:indefadjoint}
\end{equation}
The mean value of an operator $A$ in the new metric is given by
\begin{equation}
\subset\!\! A\!\!\supset_{\psi}=\frac{\ibra{\psi}A\iket{\psi}}{\ibraket{\psi}{\psi}}.
\end{equation}
If the operator is hermitian in the new metric ($A=\bar{A}$), the mean value $\subset\!\! A\!\!\supset_{\psi}$ can easily be shown to coincide with the mean $\langle A\rangle_{\psi}$ in the original metric, provided that $A=A^{\dagger}$.  Finally, for an orthonormal basis $\ket{\varphi_{j}}$, the closure relation becomes
\begin{equation}
1=\sum_{j}\ket{\varphi_{j}}\bra{\varphi_{j}}=\sum_{j}\iket{\varphi_{j}}\ibra{\varphi_{j}}M.
\end{equation}

\subsection{Construction of Hilbert Space and the Metric Operator\label{subsec:HilbertConStruct}}

As noted above in \ref{sec:whyindef}, quantizing the potentials of the Fermi Lagrangian yields a Hilbert space of states with an indefinite metric.  Following \cite{CohenTannoudjiPhotonsAndAtoms}, we will denote the adjoint of an operator $A$ as $\bar{A}$ in this metric, reserving the $A^{\dagger}$ adjoint for the transformed ``physical metric'' used in the fully covariant theory to isolate the unphysical modes.  Since we would like to perform calculations in a Hilbert space of coupled harmonic oscillators with positive-definite metric, we need to change the sign of \eqref{eq:aadcommutator} for $r=0$.  Assuming that such a metric exists, it must be related to the original indefinite metric operators by a metric operator $M$ such that
\begin{align}
Ma_{0,1,2,3}(\vec{k})M=&\; a_{0,1,2,3}(\vec{k}) & M\bar{a}_{1,2,3}(\vec{k})M=&\; a_{1,2,3}^{\dagger}(\vec{k}),
\end{align}
and
\begin{equation}
M\bar{a}_{0}(\vec{k})M=-a_{0}^{\dagger}(\vec{k}).\label{eq:MbaraM}
\end{equation}
With this transformation of the field operators, the covariant commutation relations \eqref{eq:aadcommutator} become
\begin{equation}
[a_{r}(\vec{k}),a_{s}^{\dagger}(\vec{k})]=\delta_{rs}\delta(\vec{k}-\vec{k}').\label{eq:aadtransformedcommutator}
\end{equation}
It is then straightforward to use these operators to define a well-behaved basis for the scalar polarization modes for each $\vec{k}$ in terms of the transformed operators as
\begin{equation}
|n_{0}\rangle=\frac{\adn{0}{n_{0}}}{\sqrt{n_{0}!}}\ket{0},\label{eq:scalarphysbasis}
\end{equation}
where the dependence on $\vec{k}$ is suppressed.  Because the scalar mode commutator \eqref{eq:aadtransformedcommutator} matches that of the conventional harmonic oscillator, the usual ladder operator relations apply in this basis, and \emph{all} states have positive norm ($\bra{n_{0}}n_{0}\rangle=1$).  On this basis, we can now explicitly write $M$ as \cite{CohenTannoudjiPhotonsAndAtoms}
\begin{equation}
M|n_{0}\rangle=(-1)^{n_{0}}|n_{0}\rangle.
\end{equation}
This form of $M$ can easily be shown to satisfy \eqref{eq:MbaraM} on the chosen basis, and is self-evidently hermitian in the new or ``physical'' metric.  In particular, since $\iket{\psi}\;=\ket{\psi}$ and $\ibra{\psi}=\bra{\psi}M$, we have
\begin{equation}
\ibraket{n_{0}}{n_{0}'}\;=\bra{n_{0}}M\ket{n_{0}'}=(-1)^{n_{0}'}\delta_{n_{0},n_{0}'},
\end{equation}
demonstrating that the combination of the chosen basis \eqref{eq:scalarphysbasis} with $M$ is consistent with the properties of the norm in \eqref{eq:indefnegnorm}, derived by canonical quantization of the potentials.  

A basis for the Hilbert space can be defined in the new metric as
\begin{equation}
\ket{n_{1},n_{2},n_{3},n_{0}}=\frac{\adn{1}{n_{1}}\adn{2}{n_{2}}\adn{3}{n_{3}}\adn{0}{n_{0}}}{\sqrt{n_{1}!n_{2}!n_{3}!n_{0}!}}\ket{0},
\end{equation}
although the subspace of states satisfying the modified Maxwell equations given in  \eqref{eq:smelageqmo} is necessarily smaller.  To apply the Lorenz gauge condition  \eqref{eq:lorenzgauge} to isolate the physical subspace, we must keep in mind that it is defined in the indefinite metric
\begin{equation}
\ibra{\psi} \partial_{\alpha}A^{\alpha}\iket{\psi}\; =0.\label{eq:densindeflorenzcond}
\end{equation}
Because it is not possible to form a basis in which $\partial_{\alpha}A^{\alpha}\iket{\psi}\;=0$, the Lorenz condition is typically expressed in terms of the weaker condition due to Gupta and Bleuler \cite{GuptaBleuler}
\begin{multline}
\left(\ao{3}(\vec{k})-\ao{0}(\vec{k})\right)\iket{\psi}\;=\;0 \text{, and }\\
 0=\;\ibra{\psi}\left(\ab{3}(\vec{k})-\ab{0}(\vec{k})\right).\label{eq:guptalorenz}
\end{multline}
Note that in general, expressions given in terms of operators acting on states $\iket{\psi}$ in one metric do not necessarily have the same form when expressed in terms of operators acting on the corresponding states $\ket{\psi}$ in another metric.
In the present case, however, $\ket{\psi}=\iket{\psi}$, and $M$ does not alter the annihilation operators, so $(\ao{3}-\ao{0})\iket{\psi}\;=(\ao{3}-\ao{0})\ket{\psi}$.  It is therefore convenient to work in the modified basis
\begin{equation}
\ket{n_{1},n_{2},n_{d},n_{g}}=\frac{\adn{1}{n_{1}}\adn{2}{n_{2}}\adn{d}{n_{d}}\adn{g}{n_{g}}}{\sqrt{n_{1}!n_{2}!n_{d}!n_{g}!}}\ket{0},
\end{equation}
where the $d$-photon and $g$-photon operators are given by
\begin{align}
\ao{d}=&\;\frac{i}{\sqrt{2}}(\ao{3}-\ao{0}),& \text{ and}&& \ao{g}=&\;\frac{1}{\sqrt{2}}(\ao{3}+\ao{0}),\label{eq:dgmodes}
\end{align}
which obey the usual bosonic commutation relations with respect to the physical (where the adjoint of $A$ is $A^{\dagger}$) metric.  This permits us to express the Lorenz condition \eqref{eq:guptalorenz} in the compact form $\ao{d}\ket{\psi}=0$.  

Note that although the Maxwell equations are satisfied by $\ket{\psi}$ for which $\ao{d}\ket{\psi}=\ao{d}\iket{\psi}\;=0$, the $\bra{\psi}$ which satisfy the Maxwell equations are \emph{not} necessarily those for which $\bra{\psi}\ad{d}=0$.  The Lorenz condition of Gupta and Bleuler, properly expressed in terms of the indefinite metric, is
\begin{align}
\ao{d}\iket{\psi}\;=\;&0, &\text{ and } && \ibra{\psi}\ab{d}=&\;0,\label{eq:lorenzcondindefinited}
\end{align}
where
\begin{align}\begin{split}
\ab{d}=&-\frac{i}{\sqrt{2}}(\ab{3}-\ab{0})=-i\ad{g},\\
\ab{g}=&\;\frac{1}{\sqrt{2}}(\ab{3}+\ab{0})=i\ad{d}.\label{eq:baradjointdgmodes}
\end{split}\end{align}
Thus we see that the Lorenz condition on $\bra{\psi}$ is $\ibra{\psi}\ab{d}=\bra{\psi}M(-i\ad{g})=0$.  In what follows, we will find it more convenient to use the indefinite metric to pick out the physical $\bra{\psi}$.  The physical subspace that satisfies \eqref{eq:guptalorenz} is now completely defined by \cite{CohenTannoudjiPhotonsAndAtoms}
\begin{equation}
\ket{n_{1},n_{2},0_{d},n_{g}}=\frac{\adn{1}{n_{1}}\adn{2}{n_{2}}\adn{g}{n_{g}}}{\sqrt{n_{1}!n_{2}!n_{g}!}}\ket{0}.
\end{equation}

Application of the Lorenz condition in both the indefinite metric on $\iket{\psi}$ as well the physical metric on $\ket{\psi}$ explicitly restricts one of the unphysical degrees of freedom.  At this point, we may be tempted to treat the so-called physical metric as if it were the ``real'' metric, and that expectation values calculated in the underlying indefinite metric should be judged according to whether they are sensible in the metric on $\ket{\psi}$.  Such an approach would be misguided.  If we consider only the $\ket{\psi}$ Hilbert space, then since the norm
\begin{equation}
\bra{n_{1},n_{2},0_{d},n_{g}}n_{1}',n_{2}',0_{d},n_{g}'\rangle=\delta_{n_{1},n_{1}'}\delta_{n_{2},n_{2}'}\delta_{n_{g},n_{g}'}\label{eq:physnormgmode}
\end{equation}
is positive for any $n_{g}$, it might then appear that the unphysical $g$-photon mode could yield quantum-mechanically valid observables that are nevertheless entirely decoupled from the transverse modes, and indeed decoupled from the state of any other field.  This interpretation would make it a practical necessity to trace over the $g$-modes when calculating expectation values.  This is no problem for the covariant theory, as the energy associated with each $g$-photon is zero, and there is no way for $g$-photons to couple to the transverse modes.  A trace over the unphysical modes would leave a pure state of the physically observed fields unchanged.  For the Lorentz-violating theory, the effects of a trace over such modes is potentially much more troubling, due to the existence of terms proportional to $\adn{g}{2}$ in the Hamiltonian.  This question of interpretation is immediately resolved if the observables are defined strictly according to their hermiticity in the underlying indefinite metric.  There, we find
\begin{align}
\begin{split}
&\ibraket{n_{1},n_{2},0_{d},n_{g}}{n_{1}',n_{2}',0_{d},n_{g}'}\\
=&\;\bra{n_{1},n_{2},0_{d},n_{g}}M\ket{n_{1}',n_{2}',0_{d},n_{g}'}\\
=&\;\delta_{n_{1},n_{1}'}\delta_{n_{2},n_{2}'}\delta_{n_{g},0}\delta_{n_{g}',0},
\end{split}\label{eq:normgmode}
\end{align}
since the action of $M$ on a state with $n$ $d$-photons and $m$ $g$-photons is, using the definitions \eqref{eq:dgmodes} and \eqref{eq:MbaraM},
\begin{eqnarray}
M\ket{n_{d},m_{g}}=i^{m-n}\ket{m_{d},n_{g}}.\label{eq:Monphysicalket}
\end{eqnarray}
From \eqref{eq:normgmode}, we see that the norm of any state satisfying \eqref{eq:lorenzcondindefinited} with $n_{g}>0$ must vanish, implying that such states cannot contribute to the eigenvalue of any observable operator.  This also implies that if a state $\ket{\psi}$ satisfying the Lorenz condition of Gupta and Bleuler can be written $\ket{\psi}=\ket{\psi}_{T}\otimes\ket{\phi}_{g}$; where $\ket{\psi}_{T}$ represents the state of the transverse modes, and $\ket{\phi}_{g}$ is the state of the $g$-photon mode; then the mean value of any physical observable $A$ must be
\begin{equation}
\frac{\ibra{\psi}A\iket{\psi}}{\ibraket{\psi}{\psi}}=\frac{\phantom{'}_{T}\bra{\psi}A\ket{\psi}_{T}}{\phantom{'}_{T}\bra{\psi}\psi\rangle_{T}},\label{eq:picktransverse}
\end{equation}
since $A$ can only act on the transverse degrees of freedom.  The underlying indefinite metric formally eliminates the need to trace over $g$-modes, simplifying the interpretation of both the covariant theory as well as the Lorentz-violating theory \cite{note4}.

\subsection{The Weak Lorenz Condition\label{subsec:weaklorenz}}
\begin{figure}[htbp]
\centering 
\includegraphics[width=3in]{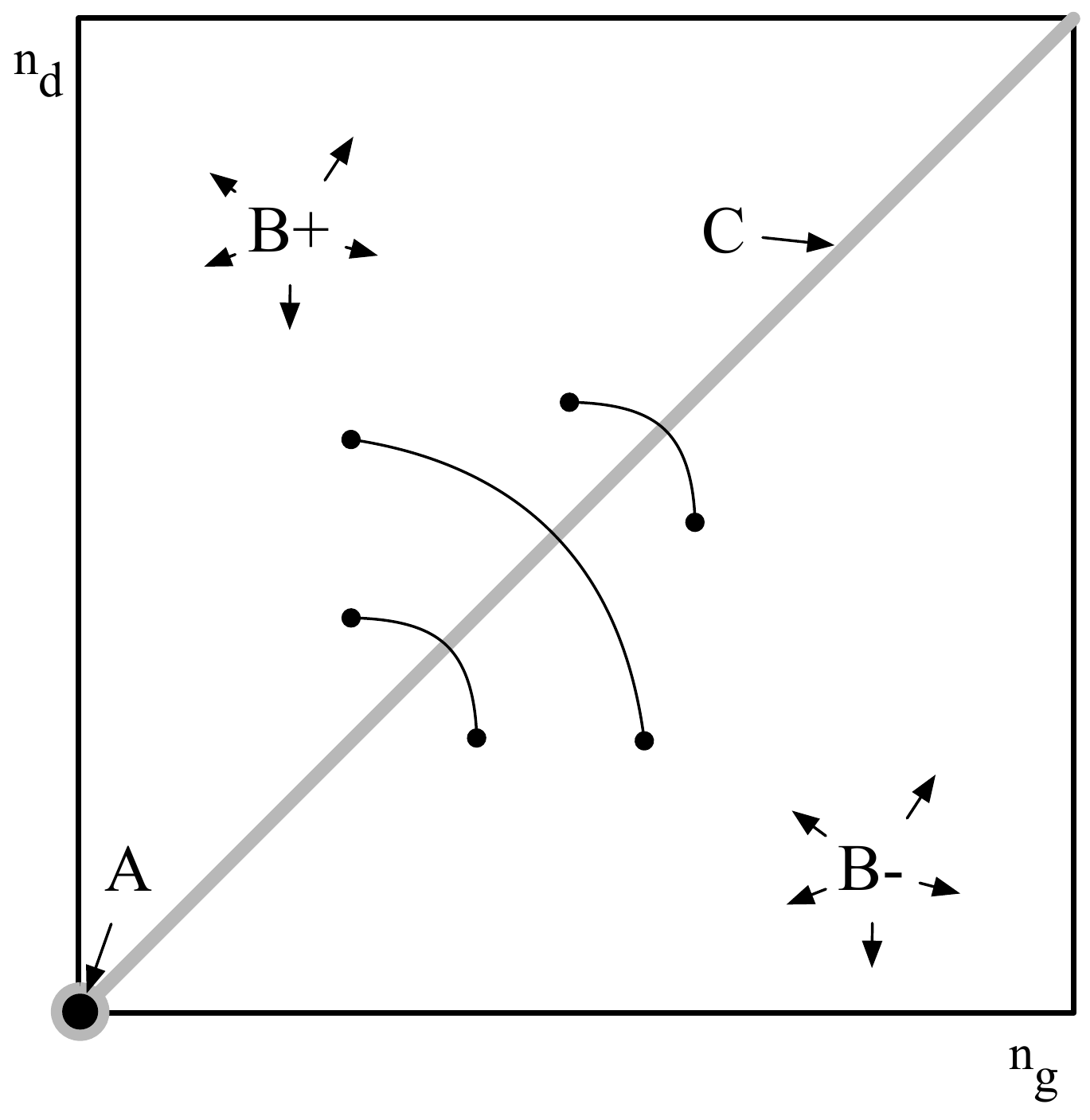}
\caption[Graphical representation of the subspace of unphysical modes]{A partition of Hilbert space into four sets according to their indefinite metric inner product.  All states within each set have a vanishing inner product with any other state in the same set.  Set A contains only the $n_{d}=n_{g}=0$ state with nonzero norm permitted by the Maxwell equations.  Set C contains all states $n_{d}=n_{g}\neq0$ that are not consistent with the Maxwell equations and have nonzero norm, while states in sets B+ and B- have varying numbers of $d$ and $g$-photons but have vanishing norm.  Each state in set B+ has a corresponding state in set B- with which it has a finite inner product.  Three such pairings are indicated by arcs.  States in sets A and C are orthogonal to states in all other sets.  A wavefunction $\iket{\psi}$ is consistent with the weak Lorenz condition \eqref{eq:weaklorenz} if it is made up of a superposition of mutually orthogonal states drawn from sets A, B+, and B-.
}
\label{fig:indefiniteHilbertMap}
\end{figure}

The preceding discussion suggests that the Lorenz condition \eqref{eq:guptalorenz} of Gupta and Bleuler may itself be stronger than is strictly necessary to satisfy \eqref{eq:densindeflorenzcond}.  We are motivated by the general form of \eqref{eq:normgmode}, which is
\begin{multline}
\ibraket{n_{1},n_{2},n_{d},n_{g}}{n_{1}',n_{2}',n_{d}',n_{g}'}\;\\
=i^{n_{g}'-n_{d}'}\delta_{n_{1},n_{1}'}\delta_{n_{2},n_{2}'}\delta_{n_{g},n_{d}'}\delta_{n_{g}',n_{d}}.\label{eq:generalnormgmode}
\end{multline}
This means that we can write down states (\eg, $\iket{n_{1},n_{2},4_{d},0_{g}}$) that do not satisfy \eqref{eq:guptalorenz}
, but which simultaneously have zero norm.  If $\ibraket{\varphi}{\varphi}\;=0$, then the contribution of $\iket{\varphi}$ to the expectation of any physical observable must also vanish, since an operator corresponding to a physical observable cannot depend or act upon the unphysical $d$ or $g$ modes.  That is, given a state $\iket{\psi}$ which is orthogonal to $\iket{\varphi}$, has nonzero norm, and which satisfies \eqref{eq:guptalorenz}, then the states $\iket{\phi_{1}}\;=\iket{\psi}$ and {$\iket{\phi_{2}}\;=c_{1}\iket{\psi}\nolinebreak[4]+c_{2}\iket{\varphi}$} are experimentally indistinguishable from one another, since for any operator $A$ corresponding to a physical observable,
\begin{align}\begin{split}
\langle A\rangle &=\frac{\ibra{\phi_{1}}A\iket{\phi_{1}}}{\ibraket{\phi_{1}}{\phi_{1}}}=\frac{\ibra{\psi}A\iket{\psi}}{\ibraket{\psi}{\psi}}\\
&=\frac{\ibra{\psi}A\iket{\psi}}{\ibraket{\psi}{\psi}}+\frac{\ibra{\varphi}A\iket{\varphi}}{\ibraket{\psi}{\psi}}=\frac{\ibra{\phi_{2}}A\iket{\phi_{2}}}{\ibraket{\phi_{2}}{\phi_{2}}}.\label{eq:dmodeindist}
\end{split}\end{align}
Note that the validity of this expression is dependent upon the orthogonality of $\iket{\psi}$ with $\iket{\varphi}$ with respect to the \emph{indefinite metric}, and not the metric suggested by \eqref{eq:physnormgmode}.  In particular, if we take $\iket{\varphi}\;=\iket{0_{1},0_{2},n_{d},0_{g}}$, then $\iket{\psi}$ must not have a $\iket{0_{1},0_{2},0_{d},n_{g}}$ component, since this would lead to a nonvanishing cross term proportional to the real part of $c_{1}c_{2}^{*}(i)^{n}$ in \eqref{eq:dmodeindist}.  A diagram of the relative orthogonality and norm of the $d$- and $g$-mode subspace for fixed $\vec{k}$ is given in Figure \ref{fig:indefiniteHilbertMap}.

If the observed field configuration in state $\iket{\phi_{1}}$ is indistinguishable from that in state $\iket{\phi_{2}}$, then since the configuration due to $\iket{\phi_{1}}$ is consistent with the (modified) Maxwell equations \eqref{eq:smelageqmo}, the field configuration represented by $\iket{\phi_{2}}$ must \emph{also} be a solution to \eqref{eq:smelageqmo}.  Thus the conventional formulation of the Lorenz gauge condition of Gupta and Bleuler is overly restrictive; it excludes states that are consistent with the Maxwell equations.  We are therefore led to restate the Lorenz condition in the less restrictive form:
\begin{equation}
\begin{split}
\text{For all } \iket{\psi} \text{ such that } \ibraket{\psi}{\psi}\neq 0: \\
\bigg(\ao{d}\iket{\psi}=0 \text{ and } \ibra{\psi}\ab{d}=0 \bigg). \label{eq:weaklorenz}
\end{split}
\end{equation}
Just as happened with respect to the $g$-photon modes in part \ref{subsec:HilbertConStruct}, the difference between the weak Lorenz condition \eqref{eq:weaklorenz} and the stronger condition of Gupta and Bleuler is relatively unimportant to the development of the fully covariant theory.  States $\iket{\varphi}$ with one or more $d$-photons such that $\ibraket{\varphi}{\varphi}\;=0$ are, like the states with one or more $g$-photons, entirely decoupled from the transverse modes as $\kf\rightarrow\nolinebreak0$.  The distinction is however critically important to the development of the Lorentz-violating theory, as the Hamiltonian \eqref{eq:DHam}, in the $\wtcal{H}_{LS}$ and $\wtcal{H}_{\pm,T,LS}$ terms, includes couplings between states that satisfy \eqref{eq:guptalorenz} and states that do not.  In what follows, we demonstrate that the Lorentz-violating Hamiltonian $\wtcal{H}_{R}$ does in fact leave the space of states that satisfy the weak Lorenz condition invariant, and therefore represents a generator of unitary time translations that is fully consistent with the modified Maxwell equations.

\subsection{Lorentz-Violating Hamiltonian in the Indefinite Metric\label{subsec:LVHIM}}
\renewcommand{\ao}[2]{\ifthenelse{\equal{#2}{}}{\ensuremath{a_{#1}}}{\ifthenelse{\equal{#2}{-k}}{\ensuremath{a_{#1}(-\vec{k})}}{\ensuremath{a_{#1}(\vec{#2})}}}}
\renewcommand{\ab}[2]{\ifthenelse{\equal{#2}{}}{\ensuremath{\overline{a}_{#1}}}{\ifthenelse{\equal{#2}{-k}}{\ensuremath{\overline{a}_{#1}(-\vec{k})}}{\ensuremath{\overline{a}_{#1}(\vec{#2})}}}}
\renewcommand{\ad}[2]{\ifthenelse{\equal{#2}{}}{\ensuremath{a^{\dagger}_{#1}}}{\ifthenelse{\equal{#2}{-k}}{\ensuremath{a^{\dagger}_{#1}(-\vec{k})}}{\ensuremath{a^{\dagger}_{#1}(\vec{#2})}}}}
\renewcommand{\abd}[2]{\ifthenelse{\equal{#2}{}}{\ensuremath{\overline{a}^{\dagger}_{#1}}}{\ifthenelse{\equal{#2}{-k}}{\ensuremath{\overline{a}^{\dagger}_{#1}(-\vec{k})}}{\ensuremath{\overline{a}^{\dagger}_{#1}(\vec{#2})}}}}

At the conclusion of part \ref{sec:Hamiltonian}, we stated that the effects of $\wtcal{H}_{LS}$, $\wtcal{H}_{+,T,LS}$, and $\wtcal{H}_{-,T,LS}$ could be ignored at leading order in $\tilde{\kappa}$.  In the limit that $\kf\rightarrow 0$, these terms pose no special problem: the $\wtcal{H}_{\pm,T,LS}$ terms vanish, and $\wtcal{H}_{LS}$ reduces to $\wtcal{H}_{LS}^{0}$, where we may explicitly make the division
\begin{equation}
\wtcal{H}_{LS}=\wtcal{H}_{LS}^{0}+\wtcal{H}_{LS}^{LV},
\end{equation}
with
\begin{align}\begin{split}
\wtcal{H}_{LS}^{0}&=\:i\hbar\omega_{k}\left(\ao{d}{k}\ab{g}{k}-\ao{g}{k}\ab{d}{k}\right.\\
&\quad\quad\left.+\ab{g}{-k}\ao{d}{-k}-\ab{d}{-k}\ao{g}{-k}\right)
\end{split}\end{align}
which becomes 
\begin{align}\begin{split}
\wtcal{H}_{LS}^{0}&=\:\hbar\omega_{k}\left(\ao{d}{k}\ad{d}{k}+\ao{g}{k}\ad{g}{k}\right.\\
&\quad\quad\left.+\ad{d}{-k}\ao{d}{-k}+\ad{g}{-k}\ao{g}{-k}\right)
\end{split}\end{align}
when expressed in the ``physical'' metric, and
\begin{align}\begin{split}
\wtcal{H}_{LS}^{LV}=&\;\;-\hbar\omega_{k}\epvkemktrepv{3}{3}{k}\times\\
&\quad\quad\quad\left(\ao{g}{k}\ab{g}{k}+\ab{d}{-k}\ao{d}{-k}\right)\\
&-i\hbar\omega_{k}\epvkemktrepv{3}{3}{k}\times\\
&\quad\quad\quad\left(\ao{g}{k}\ao{d}{-k}-\ab{d}{-k}\ab{g}{k}\right).
\end{split}\end{align}
Given the commutation relation for $[\ao{r}{k},\ab{s}{k'}]$ and the definition of $\ao{d}{k}$ and $\ao{g}{k}$, we may derive the commutators for $\ao{g}{k}$, $\ao{d}{k}$ and their adjoints:
\begin{subequations}
\begin{align}
[\ao{g}{k},\ab{g}{k'}]&=0\\
[\ao{d}{k},\ab{d}{k'}]&=0\\
[\ao{d}{k},\ab{g}{k'}]&=i\delta(\vec{k}-\vec{k'})\\
[\ao{g}{k},\ab{d}{k'}]&=i\delta(\vec{k}-\vec{k'}).
\end{align}
\end{subequations}
Using these commutation relations, it is straightforward to demonstrate that $\wtcal{H}_{LS}^{LV}$, $\wtcal{H}_{+,T,LS}$, and $\wtcal{H}_{-,T,LS}$ all commute with one another, as do the individual operators in $\wtcal{H}_{LS}^{LV}$.  To get a sense for the action of $\wtcal{H}_{LS}^{LV}$ on an arbitrary wavefunction, we must write it in terms of the ``physical'' metric, where we have defined our basis.  Using \eqref{eq:baradjointdgmodes}, we obtain
\begin{align}\begin{split}
\wtcal{H}_{LS}^{LV}=&\;\;-i\hbar\omega_{k}\epvkemktrepv{3}{3}{k}\times\\
&\quad\quad\quad\left(\ao{g}{k}\ad{d}{k}-\ad{g}{-k}\ao{d}{-k}\right)\\
&-i\hbar\omega_{k}\epvkemktrepv{3}{3}{k}\times\\
&\quad\quad\quad\left(\ao{g}{k}\ao{d}{-k}-\ad{g}{-k}\ad{d}{k}\right).
\end{split}\end{align}
Note that while $\wtcal{H}_{LS}^{LV}$ is manifestly self-adjoint with respect to the indefinite metric, it is not with respect to the physical metric.  Fortunately, the properties of the inner product are such that although $\wtcal{H}_{LS}^{LV}$ does represent a non-hermitian Hamiltonian coupling to states with different numbers of unphysical $d$- and $g$-photons, the evolution of the wavefunction with respect to physical observables (including $\wtcal{H}$) remains unitary.  As we now demonstrate, if a state $\iket{\varphi}$ is coupled by $\wtcal{H}_{LS}^{LV}$ to a state $\iket{\psi}$ with nonzero norm that also satisfies the weak Lorenz condition \eqref{eq:weaklorenz}, then $\iket{\varphi}$ must also satisfy \eqref{eq:weaklorenz}, and thus $\ibraket{\varphi}{\varphi}\;=0$.  For fixed $\vec{k}$, $\wtcal{H}_{LS}^{LV}$ can either create a $d$-photon in mode $\vec{k}$ while removing a $g$-photon from that mode, create a $g$-photon in mode $-\vec{k}$ while removing a $d$-photon from that mode, annihilate a $g$-photon from mode $\vec{k}$ along with a $d$-photon in mode $-\vec{k}$, or create a $g$-photon in mode $-\vec{k}$ along with a $d$-photon in mode $\vec{k}$.  The action of $\left(\wtcal{H}_{LS}^{LV}\right)^{N}$ on an arbitrary state $\ket{\psi}=\ket{n_{d},n_{g}}_{\vec{k}}\ket{n_{d}',n_{g}'}_{-\vec{k}}$ can yield superpositions of the states $\ket{\varphi}=\ket{m_{d},m_{g}}_{\vec{k}}\ket{m_{d}',m_{g}'}_{-\vec{k}}$, where
\begin{align}\begin{split}
m_{d}&=n_{d}+w+z\\
m_{g}&=n_{g}-w-y\\
m_{d}'&=n_{d}'-x-y\\
m_{g}'&=n_{g}'+x+z\\
N&=w+x+y+z.
\end{split}\end{align}
For $\ibraket{\varphi}{\varphi}\;\neq 0$, we must have $m_{d}=m_{g}$ and $m_{d}'=m_{g}'$, or
\begin{align}\begin{split}
m_{d}-m_{g}&=n_{d}-n_{g}+w-x+N=0\\
m_{d}'-m_{g}'&=n_{d}'-n_{g}'+w-x-N=0.\label{eq:nonzeronormcondHLS}
\end{split}\end{align}
If $\ibraket{\psi}{\psi}\;\neq 0$, then $n_{d}=n_{g}$ and $n_{d}'=n_{g}'$.  We then see that \eqref{eq:nonzeronormcondHLS} can only be satisfied for the trivial case $N=0$, and thus no power of $\wtcal{H}_{LS}^{LV}$ can couple a state $\iket{\psi}$ that satisfies the weak Lorenz condition \eqref{eq:weaklorenz} to one that does not.  Furthermore, it cannot couple two different states with nonzero norm to one another.  This means that the presence of $\wtcal{H}_{LS}^{LV}$ does not contribute to the expectation value of $\wtcal{H}$, and indeed cannot affect the expectation value of the operator for any physical observable constructed from the transverse mode operators.

We now apply a similar analysis to the $\wtcal{H}_{\pm,T,LS}=\wtcal{H}_{+,T,LS}+\wtcal{H}_{-,T,LS}$ terms.  In the physical metric, these terms take the form
\begin{widetext}
\begin{align}\begin{split}
\wtcal{H}_{\pm,T,LS}=&\;\left(\delta_{1}\ao{1}{k}+\delta_{2}\ao{2}{k}+\delta_{3}\ad{1}{-k}+\delta_{4}\ad{2}{-k}\right)\left[i\ad{d}{k}+i\ao{d}{-k}\right]\\
&+\left[\ao{g}{k}-\ad{g}{-k}\right]\left(\delta_{1}\ad{1}{k}+\delta_{2}\ad{2}{k}+\delta_{3}\ao{1}{-k}+\delta_{4}\ao{2}{-k}\right),
\end{split}\end{align}
\end{widetext}
where $\delta_{1}$, $\delta_{2}$, $\delta_{3}$ and $\delta_{4}$ are terms of order $\tilde{\kappa}$.  The action of $\left(\wtcal{H}_{\pm,T,LS}\right)^{N}$ on an arbitrary state $\ket{\psi}=\ket{n_{d},n_{g}}_{\vec{k}}\ket{n_{d}',n_{g}'}_{-\vec{k}}$ can yield superpositions of states $\ket{\varphi}=\ket{m_{d},m_{g}}_{\vec{k}}\ket{m_{d}',m_{g}'}_{-\vec{k}}$, where
\begin{align}\begin{split}
m_{d}&=n_{d}+w\\
m_{g}&=n_{g}-y\\
m_{d}'&=n_{d}'-x\\
m_{g}'&=n_{g}'+z\\
N&=w+x+y+z.
\end{split}\end{align}
Thus if $\ibraket{\psi}{\psi}\;\neq 0$, then if $\ibraket{\varphi}{\varphi}\;\neq 0$, then both
\begin{align}\begin{split}
m_{d}-m_{g}&=n_{d}-n_{g}+w-y=0\\
m_{d}'-m_{g}'&=n_{d}'-n_{g}'+N-2x-w-y=0.
\end{split}\end{align}
If $w=y$, then this is satisfied for $N=2x$, provided that $n_{d}'\geq x$ and $n_{g}\geq y$.  If $\iket{\psi}$ has nonzero norm and satisfies the weak Lorenz condition \eqref{eq:weaklorenz}, then $n_{g}=n_{d}'=0$, which in turn requires the $w=x=y=0$, and that $N=0$ if we are to have $\ibraket{\varphi}{\varphi}\;\neq 0$.  

Finally, it is interesting to consider the effect of taking the actions of both $\wtcal{H}_{LS}^{LV}$ and $\wtcal{H}_{\pm,T,LS}$ together.  We then find that the product $\left(\wtcal{H}_{LS}^{LV}\right)^{N_{1}}\left(\wtcal{H}_{\pm,T,LS}\right)^{N_{2}}$ can couple $\ket{\psi}=\ket{n_{d},n_{g}}_{\vec{k}}\ket{n_{d}',n_{g}'}_{-\vec{k}}$ to $\ket{\varphi}=\ket{m_{d},m_{g}}_{\vec{k}}\ket{m_{d}',m_{g}'}_{-\vec{k}}$ provided that
\begin{align}\begin{split}
m_{d}&=n_{d}+w_{1}+z_{1}+w_{2}\\
m_{g}&=n_{g}-w_{1}-y_{1}-y_{2}\\
m_{d}'&=n_{d}'-x_{1}-y_{1}-x_{2}\\
m_{g}'&=n_{g}'+x_{1}+z_{1}+z_{2}\\
N_{1}&=w_{1}+x_{1}+y_{1}+z_{1}\\
N_{2}&=w_{2}+x_{2}+y_{2}+z_{2}.
\end{split}\end{align}
If $\ket{\varphi}$ has a nonzero norm, then we must have
\begin{align}\begin{split}
m_{d}-m_{g}=&\;n_{d}-n_{g}+w_{1}-x_{1}+N_{1}+w_{2}-y_{2}=0\\
m_{d}'-m_{g}'=&\;n_{d}'-n_{g}'+w_{1}-x_{1}\\
&-N_{1}+N_{2}-2x_{2}-w_{2}-y_{2}=0.
\end{split}\end{align}
If $\iket{\psi}$ satisfies \eqref{eq:weaklorenz}, then $w_{1}=y_{1}=y_{2}=y_{1}=x_{1}=x_{2}=0$, and the above reduces to
\begin{align}\begin{split}
m_{d}-m_{g}&=n_{d}-n_{g}+N_{1}+w_{2}=0\\
m_{d}'-m_{g}'&=n_{d}'-n_{g}'-N_{1}+N_{2}-w_{2}=0,
\end{split}\end{align}
which cannot be satisfied for any $N_{1}>0$ or $N_{2}>0$.  Taking the subspace $S_{LV}$ as that generated by $\wtcal{H}_{R}$ on the subspace $S$ of states with no $d$- or $g$-mode excitations, we may now say that every $\iket{\psi}\;\in S_{LV}$ satisfies the weak Lorenz condition \eqref{eq:weaklorenz}.  This means that $\wtcal{H}_{R}$ leaves the space of solutions of the modified Maxwell equations \eqref{eq:smelageqmo} invariant.  Furthermore, we have shown that the apparently non-hermitian form of $\wtcal{H}_{LS}^{LV}$ and $\wtcal{H}_{\pm,T,LS}$ in terms of the physical metric operators does not lead to non-unitary evolution in time, since all states coupled by such terms have vanishing norm.

\section{Effects on Transverse Mode Couplings\label{sec:quanthameffects}}
Although this work focuses on the free-field evolution, it is worthwhile to consider the form of the transverse potentials when expressed in terms of the free-field eigenmodes.  From \eqref{eq:kAaad}, we find that the transverse components of the potential
\begin{eqnarray}
\kAd{\perp,1}{k}&=&\sqrt{\frac{\hbar c^{2}}{2\omega_{k}}}\left(\ao{1}{k}+\ab{1}{-k}\right)\\
\kAd{\perp,2}{k}&=&\sqrt{\frac{\hbar c^{2}}{2\omega_{k}}}\left(\ao{2}{k}+\ab{2}{-k}\right)
\end{eqnarray}
become
\begin{align}\begin{split}
e^{\Xi_{1}+\Xi_{2}}\,\kAd{\perp,1}{k}\,&e^{-\Xi_{1}-\Xi_{2}}=\\
& \left(1-\delta_{1}\right)\kAd{\perp,1}{k}-\delta_{2}\kAd{\perp,2}{k}
\end{split}\\ \begin{split}
e^{\Xi_{1}+\Xi_{2}}\,\kAd{\perp,2}{k}\,&e^{-\Xi_{1}-\Xi_{2}}=\\
&\left(1+\delta_{1}\right)\kAd{\perp,2}{k}-\delta_{2}\kAd{\perp,1}{k}
\end{split}\end{align}
where
\begin{equation}
\begin{split}
\delta_{1}=&\frac{1}{4}\epvkemktrepv{1}{1}{k}\\
&-\frac{1}{4}\epvkemktrepv{2}{2}{k}\\
\delta_{2}=&\frac{1}{2}\epvkemktrepv{1}{2}{k}.
\end{split}
\end{equation}
This has consequences for the interaction of the field eigenstates with a charged current $j_{\mu}$, which is given, up to a normalization constant, by
\begin{equation}
V_{\rm int.}=-\int d^{3}k \sum_{r} \epsilon_{r}^{\mu}(\vec{k})\left(j_{\mu}(\vec{k})\overline{\mathcal{A}_{r}}(\vec{k})+j_{\mu}^{*}(\vec{k})\kAd{r}{k}\right).
\end{equation}
For a wave propagating with wave vector $k_{z}$ in the $+\hat{z}$ direction, with the two orthogonal polarizations lying respectively along the $\hat{x}$ and $\hat{y}$ axes, this becomes, after applying the above similarity transform to the fields
\begin{multline}
V_{\rm int.}(k_{z}) =  -\left[\left(\tfrac{4-\kem^{xx}+\kem^{yy}}{4}\right)j_{1}(k_{z})-\tfrac{\kem^{12}}{2}j_{2}(k_{z})\right]\overline{\mathcal{A}_{1}}(k_{z})\\
 -\left[\left(\tfrac{4+\kem^{xx}-\kem^{yy}}{4}\right)j_{2}(k_{z})-\tfrac{\kem^{12}}{2}j_{1}(k_{z})\right]\overline{\mathcal{A}_{2}}(k_{z})\\
-\left[\left(\tfrac{4-\kem^{xx}+\kem^{yy}}{4}\right)j_{1}(k_{z})^{*}-\tfrac{\kem^{12}}{2}j_{2}(k_{z})^{*}\right]{\mathcal{A}_{1}}(k_{z})\\
 -\left[\left(\tfrac{4+\kem^{xx}-\kem^{yy}}{4}\right)j_{2}(k_{z})^{*}-\tfrac{\kem^{12}}{2}j_{1}(k_{z})^{*}\right]{\mathcal{A}_{2}}(k_{z})\label{eq:effAint}
\end{multline}

A more complete treatment of Lorentz-violating QED coupled to matter would further transform the coupling between matter and the field's longitudinal and scalar degrees of freedom, 
reproducing the anisotropic Coulomb potential first derived in~\cite{Bailey:2004}.  Nevertheless, the interaction term given in Eq.~\eqref{eq:effAint} 
is sufficient to show that to leading order in $\tilde{\kappa}$, the strength of the interaction between a propagating wave and an isotropic charge distribution depends upon the orientation of the wave's transverse polarization.  Thus although the SME parameters under consideration do not cause the vacuum to become birefringent, they cause light to interact with charges in a birefringent manner.  
This means that a Michelson-Morley test could be performed by searching for frame-dependence in the refractive index for two orthogonally polarized optical modes propagating within a single dielectric cavity, rather than requiring the use of two separate resonators.  This is consistent with recent analyses of the classical~\cite{Kostelecky:2009}, and coordinate-transformed semi-classical~\cite{Mueller:2005} theory.  An extension of the derivation presented here incorporating the interaction of the potentials with charged particles would likely aid in such analyses, and will be the subject of future work.

\section{Conclusion}

We have demonstrated that the free electromagnetic field in the non-birefringent limit of the minimal SME can be quantized, possesses a stable vacuum state, and that its evolution is Hermitian on the set of states which satisfy the modified Maxwell equations of motion.  We have shown that the theory requires the use of a weak Lorenz gauge condition, which allows the set of physical states $S_{LV}$ to include states with vanishing norm.  The inclusion of these states is necessary for the Hermiticity of the Lorentz-symmetry breaking theory.  

The single-photon Lorentz-symmetry violating eigenstates have been shown to satisfy the same dispersion relation as do the solutions to the classical field equations.  This provides a firmer theoretical basis on which to  analyze atom-photon scattering experiments~\cite{ModernIvesStillwell} in coordinates such that electromagnetism breaks Lorentz symmetry.  This also raises the prospect of developing consistent experimental tests of Lorentz symmetry at the few or single-photon level.  As previously demonstrated in the classical limit~\cite{Kostelecky:2009}, the quantized fields generally couple to matter anisotropically.  Although the full form of the quantized matter-coupled theory has not been developed, this suggests that precise cavity QED experiments may be able to perform unique tests of Lorentz invariance in the context of a fully quantum system.  Finally, we find that the number of photons in a given field eigenstate is not, in general, a conserved quantity, and can change under boost transformations.  While determination of the degree to which this leads to observable effects awaits the development of the full charge-coupled theory, this suggests that recent developments in relativistic quantum information theory may have application to tests of Lorentz invariance~\cite{Peres:2004}.

\acknowledgments
We thank Don Colladay, V. Alan Kosteleck\'y and Matthew Mewes for useful discussions.

\begin{widetext}
\appendix*
\section{$\kf$ Identities\label{sec:riemannsymmetries}}
Since $k_{F}$ has the symmetries of the Riemann tensor, we know that
\begin{equation}
\kf_{\kappa \lambda \mu \nu}=\kf_{\mu \nu \kappa \lambda}=-\kf_{\lambda \kappa \mu \nu}=-\kf_{\kappa \lambda \nu \mu}
\end{equation}
and
\begin{equation}
\kf_{\kappa\lambda \mu \nu}+\kf_{\kappa \nu \lambda \mu}+\kf_{\kappa \mu \nu\lambda}=0.
\end{equation}
This means that given a set of four $4$-vectors
\begin{eqnarray}
v&=&(v^{0},\vec{v})\\
x&=&(x^{0},\vec{x})\\
y&=&(y^{0},\vec{y})\\
z&=&(z^{0},\vec{z}),
\end{eqnarray}
then the product (summed over repeated indexes) $\kf_{\kappa\lambda\mu\nu}w^{\kappa}x^{\lambda}y^{\mu}z^{\nu}$ can be written as
\begin{eqnarray*}
(k_{F})_{\kappa\lambda\mu\nu}w^{\kappa}x^{\lambda}y^{\mu}z^{\nu}&=&
\kf_{0j0k}\left(\wxyz{0}{j}{0}{k}+\wxyz{j}{0}{k}{0}-\wxyz{0}{k}{j}{0}-\wxyz{j}{0}{0}{k}\right)\\
&&+\kf_{jl 0k}\left(\wxyz{0}{k}{j}{l}-\wxyz{j}{l}{k}{0}+\wxyz{j}{l}{0}{k}-\wxyz{k}{0}{j}{l}\right)\\
&&+\kf_{jl km}\wxyz{j}{l}{k}{m}
\end{eqnarray*}
In terms of $\kem$, $\kop$, and $\ktr$, this becomes
\begin{align}\begin{split}
(k_{F})_{\kappa\lambda\mu\nu}w^{\kappa}x^{\lambda}y^{\mu}z^{\nu}=&-\frac{1}{2}\left(\left[w^{0}\vec{x}-\vec{w}x^{0}\right]\cdot\left(\kem+I\ktr\right)\cdot\left[y^{0}\vec{z}-\vec{y}z^{0}\right]\right)\\
&\:\:-\frac{1}{2}\left(\left[w^{0}\vec{x}-\vec{w}x^{0}\right]\cdot\kop\cdot\left[\vec{y}\times\vec{z}\right]+\left[y^{0}\vec{z}-\vec{y}z^{0}\right]\cdot\kop\cdot\left[\vec{w}\times\vec{x}\right]\right)\\
&\:\:-\frac{1}{2}\left(\left[\vec{w}\times\vec{x}\right]\cdot\left(\kem+I\ktr\right)\cdot\left[\vec{y}\times\vec{z}\right]\right).\label{eq:kfVectorIdent}
\end{split}\end{align}
\end{widetext}



\begin{thebibliography}{99}
\bibitem{note2}
To be consistent with the notation of~\cite{MandlAndShaw}, we have departed from our usual convention that reserves roman indices for 3-vectors, as $r$ now denotes the scalar ($r=0$), transverse ($r=1,2$) and longitudinal ($r=3$) modes for a given wavevector $\vec{k}$, rather than the components of a 3 or 4-vector.
\bibitem{note3}
Note that this statement applies to the classical as well as the quantum theory.  Differences in the derivation of the classical covariant field representation as compared to their quantum representation arise according to how the Lorenz gauge condition is applied.
\bibitem{note4}
We could have arrived at an expression similar to \eqref{eq:normgmode}, and thus derived \eqref{eq:picktransverse} purely in terms of the $\ket{\psi}$ metric, using the properly transformed adjoint of the Lorenz condition $\bra{\psi}M(-ia^{\dagger}_{g})=0$.  This would show that $\bra{n_{g}\neq0}$ does not belong to the subspace satisfying \eqref{eq:guptalorenz} on the larger Hilbert space.  The problem is somewhat easier to address both mathematically and conceptually in the indefinite metric.
\end{thebibliography}
\end{document}